\documentclass[11pt]{article}
\usepackage{graphicx}
\usepackage{epsfig}
\usepackage{bbm}

\newcommand{\be}{\begin{eqnarray}}
\newcommand{\ee}{\end{eqnarray}}

\def\nue{{\nu_e}}
\def\anue{{\bar\nu_e}}
\def\numu{{\nu_{\mu}}}
\def\anumu{{\bar\nu_{\mu}}}

\newcommand{\ms}{\Delta m^2_{21}}
\newcommand{\ma}{\Delta m^2_{31}}
\newcommand{\meff}{\Delta m^2_{\rm eff}}

\newcommand{\stch}{\sin^2 2\theta_{13}}
\newcommand{\sa}{\sin^2 \theta_{23}}

\newcommand{\dcp}{\delta_{CP}}

\def\gtap{\ \raisebox{-.4ex}{\rlap{$\sim$}} \raisebox{.4ex}{$>$}\ }

\def\gs{\mathrel{
   \rlap{\raise 0.511ex \hbox{$>$}}{\lower 0.511ex \hbox{$\sim$}}}}
\def\ls{\mathrel{
   \rlap{\raise 0.511ex \hbox{$<$}}{\lower 0.511ex \hbox{$\sim$}}}}
\newcommand{\bea}{\begin{equation} \begin{array}{c}}
\newcommand{\bead}{\begin{equation} \begin{array}{cccc}}
\newcommand{\eea}{ \end{array} \end{equation}}

\begin{document}

\title{\bf Measuring the Mass Hierarchy with Muon and Hadron Events in 
Atmospheric Neutrino Experiments
}
\author{
\\
Anushree Ghosh$^a$\,\thanks{email:\tt anushree@hri.res.in}~,~~
Sandhya Choubey$^a$\,\thanks{email: \tt sandhya@hri.res.in}~,~~
\\\\
{\normalsize \it$^{a}$Harish-Chandra Research Institute, Chhatnag Road, Jhunsi, Allahabad 211 019, India}\\ \\
}

\maketitle
\begin{abstract}\noindent

Neutrino mass hierarchy can be measured in atmospheric neutrino experiments 
through the observation of earth matter effects. Magnetized iron calorimeters have 
been shown to be good in this regard due to their charge identification capabilities.
The charged current interaction of $\numu$ in this detector, produces a muon track and 
a hadron shower. The direction of the muon track can be 
measured very accurately. 
We show the improvement expected in the reach of this 
class of experiments to the neutrino mass hierarchy, as 
we improve the muon energy resolution and the muon 
reconstruction efficiency. We next propose to include the hadron events in the 
analysis, by tagging them with the zenith angle of the corresponding muon and binning 
the hadron data first in energy and then in zenith angle. To the best of our knowledge this way of 
performing the analysis of the atmospheric neutrino data has not be considered before. 
We show that the hadron events increase 
the mass hierarchy sensitivity of the experiment. Finally, we show the expected mass hierarchy 
sensitivity in terms of the 
reconstructed neutrino energy and zenith angle. 
We show how 
the detector resolutions spoil the earth matter effects in the 
neutrino channel and argue why the sensitivity obtained 
from the neutrino analysis cannot be significantly better than 
that obtained from the analysis using muon data alone. 
As a result, the best mass hierarchy sensitivity is 
obtained when we add the contribution of the muon and the hadron data.
For $\sin^22\theta_{13}=0.1$, $\sin^2\theta_{23}=0.5$, 
a muon energy resolution of 2\%, reconstruction efficiency of 80\% and 
exposure of $50\times 10$ kton-year, 
we could get up to $4.5\sigma$ signal for the mass hierarchy from combining the muon 
and hadron data. The signal will go up when the atmospheric data is combined with data 
from other existing experiments, particularly NO$\nu$A.

\end{abstract}

\newpage

\section{Introduction}

The latest set of data from the reactor-based experiments Daya Bay \cite{dbth13}, RENO \cite{renoth13} and 
Double Chooz \cite{dcth13} have confirmed beyond all doubt 
what the accelerator-based experiments T2K \cite{t2kth13} and MINOS \cite{minosth13} 
had earlier indicated -- the value of the neutrino mixing parameter $\stch \simeq 0.1$. The 
implications of this discovery has far reaching implications in neutrino physics. 
On the phenomenological front, this opens up the possibility for the determination of the 
next two missing links in neutrino oscillation physics, {\it viz.}, (i) the discovery of 
CP violation in the lepton sector, and (ii) the sign of $\Delta m_{31}^2$, {\it aka}, the 
neutrino mass hierarchy. Next generation neutrino oscillation experiments are being 
proposed to probe these two remaining issues. The optimal design of the next 
set of neutrino oscillation experiments will depend on how well they can measure 
CP violation and neutrino mass hierarchy, given that now we know that $\theta_{13}$ is
much larger than previously thought. The fact that the sensitivity of the proposed experiment 
to one of these parameters could be severely restricted by the uncertainty on the other parameter 
makes the designing of the experiments all the more challenging. \\

Of the two unknowns mentioned above, measurement of CP violation is trickier for a variety of 
reasons.  CP violation in neutrino oscillations is necessarily a 
sub-leading effect and is expected to be be in the $\theta_{13}$-driven appearance channel,
$P_{\mu e}$. 
Therefore, it is necessary to have a non-zero $\theta_{13}$ for the measurement of the CP phase. 
However, the fact that $\theta_{13}$ has turned to be relatively 
large might prove counter-productive for the CP 
violation searches. The reason mainly being  that the CP violation effects in the appearance 
channel become further sub-dominant compared to the main CP independent $\theta_{13}$ 
driven oscillations for such large values of $\theta_{13}$. This makes the latter an irreducible 
background, decreasing thereby the sensitivity of the experiment 
for CP searches \cite{minakata2012,efm2012}. The uncertainty regarding the neutrino mass 
hierarchy introduces another limitation on these experiments through the $\dcp-$mass hierarchy 
parameter degeneracy, deteriorating  further the sensitivity of the experiment to the 
CP phase. \\

Measurement of the neutrino mass hierarchy on the other hand certainly becomes easier as 
the true value of $\theta_{13}$ increases. This parameter is expected to be measured using earth 
matter effects in neutrino oscillations. The earth matter effect increases monotonically with the value of 
$\theta_{13}$, making their detection in terrestrial experiments easier for larger $\theta_{13}$. 
The atmospheric neutrino experiments in this regard could play a crucial role in the 
field of neutrino physics in the future. The possibility of measuring the neutrino mass 
hierarchy in atmospheric neutrino experiments have been considered in details in the 
literature \cite{petcov2002}-\cite{bs2012}. Upcoming detectors for atmospheric neutrino include 
the magnetized Iron CALorimeter detector at the India-based Neutrino Observatory 
(ICAL@INO) \cite{ino}, the megaton-class water Cherenkov detectors such those proposed for the 
Hyper-Kamiokande project \cite{hk} and as the far detector of the CERN-Frejus long baseline 
experiment \cite{laguna}, the large liquid argon detectors \cite{lar}, as well as the 
giant ice detector PINGU \cite{pingu}. 
\\

Recently, we considered the sensitivity 
of the ICAL@INO experiment to the measurement of the neutrino mass hierarchy \cite{mh}
and the atmospheric neutrino oscillation parameters \cite{precision}. 
These studies were performed as a part of the on-going efforts of the INO collaboration 
towards the simulation of the ICAL detector and 
the analysis of the simulated data to determine the physics reach of the ICAL@INO 
experiment. When combined with 
expected data from all the current accelerator and reactor experiments, 
Daya Bay, RENO, Double Chooz, T2K and NO$\nu$A, $50\times 10$ kton-year 
of data from ICAL@INO was shown to give a mass hierarchy sensitivity of $2.3\sigma-5.7\sigma$, 
depending on the true value of $\stch$, $\sa$ and $\dcp$.  However, note that 
these analyses were based on using only the muon sample of the simulated INO 
data. The full detector response to muons obtained from detailed Geant-based simulations 
of the ICAL detector \cite{inomuon} were used in these studies and the sensitivity of the 
experiment to the neutrino oscillation parameters estimated. In particular, 
the muon energy resolution was seen to be around $\sigma_{E_\mu}/E_\mu=10-15\%$, 
below what has been found from simulations of other similar detectors such as 
MINOS and MONOLITH \cite{monolithproposal}. 
However, the simulations of 
the ICAL detector is on-going and the results obtained on the detector response to muons 
is expected to evolve along with the improvement in the detector simulation code as well as 
the analysis. Amongst other things, the reconstruction algorithm is being improved upon, 
which would allow for better reconstruction of the muon track which could lead to better 
energy resolution of the muons. In addition, ICAL will also be somewhat sensitive to the 
associated hadron(s) which produce(s) a shower. The detector response to the hadron 
shower and in particular the hadron energy resolution was recently studied by the INO 
collaboration and presented in \cite{inohadron}. 
\\

In this paper we study how much the mass hierarchy sensitivity of the ICAL@INO, or 
any other similar experiment, could improve, by improving the energy and angle 
resolution of the detector, as well as its particle 
reconstruction efficiency. We start with the muon event analysis 
and study how much the mass hierarchy sensitivity 
would improve with the energy resolution of the muons. Since the angular resolution 
of the muons obtained from the recent simulation results by the INO 
collaboration is found to be extremely good, we fix the angular resolution to 
be $\sigma_{\Theta_\mu}=0.01$ in $\cos\Theta_\mu$, a value that is consistent with the 
simulation results \cite{inomuon}, and vary only the energy resolution. We also study the 
impact of increasing the reconstruction efficiency of the muons, though this is more 
mundane as the reconstruction efficiency merely increases the overall statistics of the 
muons. 
\\

We next include the hadron events in the study and present 
results of a statistical analysis where muon and hadron events are included separately 
in a combined $\chi^2$ function. 
To the best of our knowledge, this way of analyzing the 
muon and the hadron data from atmospheric neutrino events 
has never been considered before. 
We propose a unique way of tagging the hadron events. We tag every
hadron event with the zenith angle of the corresponding muon produced in 
the charged current interaction of the neutrino. We collect all 
such hadrons in every muon zenith angle bin. This hadron sample can then
be further binned in hadron energy as well as hadron zenith angle. 
We include these hadron events in the statistical analysis along with the muons and 
show that the mass hierarchy sensitivity of the experiment improves reasonably 
with the inclusion of these events.
\\

Most analyses of the atmospheric neutrino experiments are in terms of the reconstructed 
neutrino energy and zenith angle. 
The neutrino energy and 
zenith angle can be reconstructed 
from the measured energy and zenith angle of the muon and hadron events. 
The dependence of the mass hierarchy sensitivity to the reconstructed neutrino energy and 
angle resolution functions has been studied in somewhat details in the literature 
\cite{petcov2005,gandhi2007}.
Reconstructing the neutrino energy in magnetized iron detectors 
can be done without too much difficulty by adding the measured muon energy with 
the measured hadron energy. While the final simulation results from the INO collaboration 
are still awaited, the MINOS simulations have yielded an energy resolution 
$\sigma_{E_\nu}/E_\nu \simeq 15\%$ while the MONOLITH
proposal quotes $\sigma_{E_\nu}/E_\nu \simeq 20\%$ \cite{monolithproposal}. 
On the other hand, the neutrino zenith angle will have to be extracted either 
by reconstructing the 
neutrino momentum from combining the 
muon and hadron momenta, or from simulating the angle between the 
neutrino and muon direction. 
In this paper, we will present results on the neutrino analysis 
for two cases. In first one we will bin the data in neutrino energy and neutrino zenith angle bins. 
We will use the zenith angle resolution function for the neutrinos quoted in the MONOLITH 
proposal \cite{monolithproposal}, and show the mass hierarchy sensitivity results as a function of the neutrino 
energy resolution. For the next case we will bin the data in neutrino energy and muon zenith 
angle bins.
For this case we will use a fixed muon 
zenith angle resolution function which agrees with 
the results of the INO muon simulations. We will compare the sensitivity results obtained 
from the two cases. 
 \\
 
 A discussion on marginalization of the $\chi^2$ over all the oscillation parameters in the 
 fit is in order. It is well known that marginalization over the oscillation parameters 
 does reduce the mass hierarchy sensitivity when the atmospheric data is taken alone. 
 However, we had shown in \cite{mh} that when we do a combined fit of the atmospheric 
 neutrino data along with the data from the other accelerator and reactor experiments, 
 Daya Bay, RENO, Double Chooz, T2K and NO$\nu$A, then the $\Delta \chi^2$ 
 remains the same as that obtained by keeping all parameters fixed at their 
 assumed true values. Since data from these accelerator and reactor experiments will 
 anyway be available by the time we get the atmospheric neutrino data from any 
 magnetized iron detector, it is pertinent is always add them in the fit. This is what was 
 done in \cite{mh}. In this paper, since we wish to only study how to optimize the 
 analysis of the atmospheric neutrino data for maximum mass hierarchy sensitivity, 
 we do not explicitly include the accelerator and reactor data. However, since their 
 major role in the global analysis is to keep best-fit $|\ma|$, $\sa$ and $\stch$ very close to 
 their assumed true values, 
 we incorporate this feature by keeping these parameters fixed in the fit. The main 
 purpose for this simplification is to save computation time as the main physics 
 impact is incorporated anyway. The accelerator experiment NO$\nu$A also 
 gives a contribution to the neutrino mass hierarchy. But it was shown in \cite{mh,bs2012} 
that this does not lead to any synergy between the atmospheric 
neutrino data and NO$\nu$A data. Therefore, the additional contribution from NO$\nu$A 
 (as a function of the true value of $\dcp$) 
can simply be added to the $\Delta \chi^2$ obtained from the atmospheric neutrino data.
\\
  
The paper is organized as follows. We start with a description of the earth matter effects in 
section \ref{sec:prob}. In section 3 we discuss the events in terms of the neutrino energy spectrum, 
muon energy spectrum and hadron energy spectrum. We begin our main results part in 
section 4 where we give the mass hierarchy sensitivity when only the muon data is 
included and show the effect of the 
muon energy resolution and 
reconstruction efficiency on the reach of the experiment in ruling out the wrong hierarchy. 
In section 5 we should how the mass hierarchy sensitivity improves when we include the 
independent hadron events in addition to the muon events in the analysis. Section 6 
gives the comparative results when we do the analysis in terms of the neutrino energy and neutrino 
angle as well as in terms of neutrino energy and muon zenith angle. We end the paper in 
section 7 with our conclusions. 
\\


\section{Earth Matter Effects in Oscillation Probabilities}
\label{sec:prob}

Atmospheric neutrinos (and antineutrinos) 
are produced in both $\numu$ and $\nue$ (and $\anumu$ and $\anue$) 
flavors, with a flavor ratio of roughly 
$\phi_{\numu}/\phi_{\nue} \sim 2$ at sub-GeV energies. This ratio increases 
with the neutrino energy.  The neutrinos, on their way from their point of production in 
the atmosphere to the detector, undergo flavor oscillations. On arrival at the detector, they 
produce the corresponding charged lepton through charged current interaction on 
nucleons. Since the oscillated atmospheric neutrino ``beam" is a combination of 
all three flavors, they produce electrons, muons, as well as tau leptons in the detector. 
Magnetized iron calorimeters such ICAL@INO 
can only efficiently detect the muons, and are hence 
sensitive to only muon type neutrinos. However, being magnetized, this kind of 
detector will be 
able to identify the charge of the muon and hence will be able to separate the 
$\numu$ signal from the $\anumu$ signal very efficiently. 
\\

Since the oscillated muon type neutrinos arriving at the detector are a combination of the 
survived $\numu$ produced in the atmosphere and the flavor oscillated $\numu$ 
coming from $\nue$ produced in the atmosphere, 
the oscillation probability channels relevant for atmospheric muon neutrinos are 
the survival probability $P_{\mu\mu}$ and transition probability $P_{e\mu}$. If 
for the sake of simplicity of discussion
we take $\ms=0$, then for the oscillation probabilities can be written as
\be
\nonumber
P_{\mu\mu}^{approx} =  1 \!\!\!\!\!\!\!\!\!\!&&
-\sin^2\theta_{13}^M\sin^22\theta_{23} \sin^2\frac{[(\ma + A)-(\ma)^M]L}{8E_\nu}
\\ \nonumber
&& - \cos^2\theta_{13}^M\sin^22\theta_{23} \sin^2\frac{[(\ma + A)+(\ma)^M]L}{8E_\nu}
\\
&&-\sin^22\theta_{13}^M\sin^4\theta_{23} \sin^2\frac{(\ma)^M L}{4E_\nu}
\,,
\label{eq:pmm}
\ee
and
\be
P_{e\mu}^{approx} = \sin^22\theta_{13}^M\sin^2\theta_{23} \sin^2\frac{(\ma)^M L}{4E_\nu}
\,,
\label{eq:pem}
\ee
where $A=2\sqrt{2}G_F N_e E_\nu$ is the matter potential in earth, $N_e$ being the electron density 
inside earth and 
$E_\nu$ the neutrino energy. The quantities 
$(\ma)^M$ and $\theta_{13}^M$ are 
the mass squared difference and mixing angle in constant density matter and are given as
\be
(\ma)^M = \bigg(( \ma\cos2\theta_{13}-A)^2 + \ma \sin^22\theta_{13}\bigg )^{1/2}
\,,
\label{eq:delmatter}
\ee
\be
\sin^22\theta_{13}^M = \frac{\ma \sin^22\theta_{13}}{\bigg((\ma\cos2\theta_{13}-A)^2 + \ma \sin^22\theta_{13}\bigg )}
\,.
\label{eq:th13matter}
\ee
We can see the role of $\theta_{13}$-driven  
earth matter effects and the neutrino mass hierarchy through these expressions. For 
$\theta_{13}=0$, we get $(\ma)^M = (\ma-A) $ and $\theta_{13}^M=0$ 
from Eqs. (\ref{eq:delmatter}) and (\ref{eq:th13matter}). Plugging these values in 
Eqs. (\ref{eq:pmm}) and (\ref{eq:pem}) we find that we 
would have no earth matter effects in neutrino oscillations for $\theta_{13}=0$. In particular,  
we can see that $P_{e\mu}^{approx}=0$ and 
$P_{\mu\mu}^{approx}$ is the same for both normal ($\ma >0$) 
as well as inverted hierarchy ($\ma <0$). However, for 
non-zero $\theta_{13}$ we get a difference in $P_{e\mu}^{approx}$ as well as 
$P_{\mu\mu}^{approx}$ between $\ma > 0$ and $\ma < 0$ due the earth matter effects. 
This difference can be used to distinguish between the normal and inverted 
neutrino mass hierarchy. 
\\

In the discussion above we had put $\ms=0$ and used a constant density for the 
matter for simplicity of the discussion. 
In all numerical 
results presented in this paper we use the full three-generation oscillation probability 
calculated using the 24 layer PREM profile for the earth matter density \cite{prem}. 
For $\ms \neq 0$, it turns out that the survival probability $P_{\mu\mu}$ becomes 
different for $\ma >0$ and $\ma < 0$, even for $\theta_{13}=0$. 
This aspect has been discussed in detail in \cite{gandhizero,meff}. 
It was shown that one could define an effective mass 
\be
\meff = \ma - (\cos^2\theta_{12} - \cos\delta_{CP}\sin\theta_{13}\sin2\theta_{12}\tan\theta_{23})\ms
\,,
\label{eq:meff}
\ee
such that this issue could be alleviated. 
It was shown that with this if one defined normal hierarchy as $\meff >0$ and 
inverted hierarchy as $\meff <0$, then 
$P_{\mu\mu}$ is the same for normal and inverted 
hierarchies when $\theta_{13}=0$. Though this complication in the definition of the 
mass hierarchy does not 
make much difference 
to the final neutrino mass hierarchy sensitivity results when one marginalizes correctly over $|\ma|$ 
(for a very recent discussion see \cite{mh}), we will continue to use this definition for the 
normal and inverted hierarchy, especially since in this paper we will not marginalize our 
$\chi^2$ function over the oscillation parameters, as discussed in the Introduction section.  
\\

\begin{figure}[h]
\centering
\includegraphics[width=0.495\textwidth]{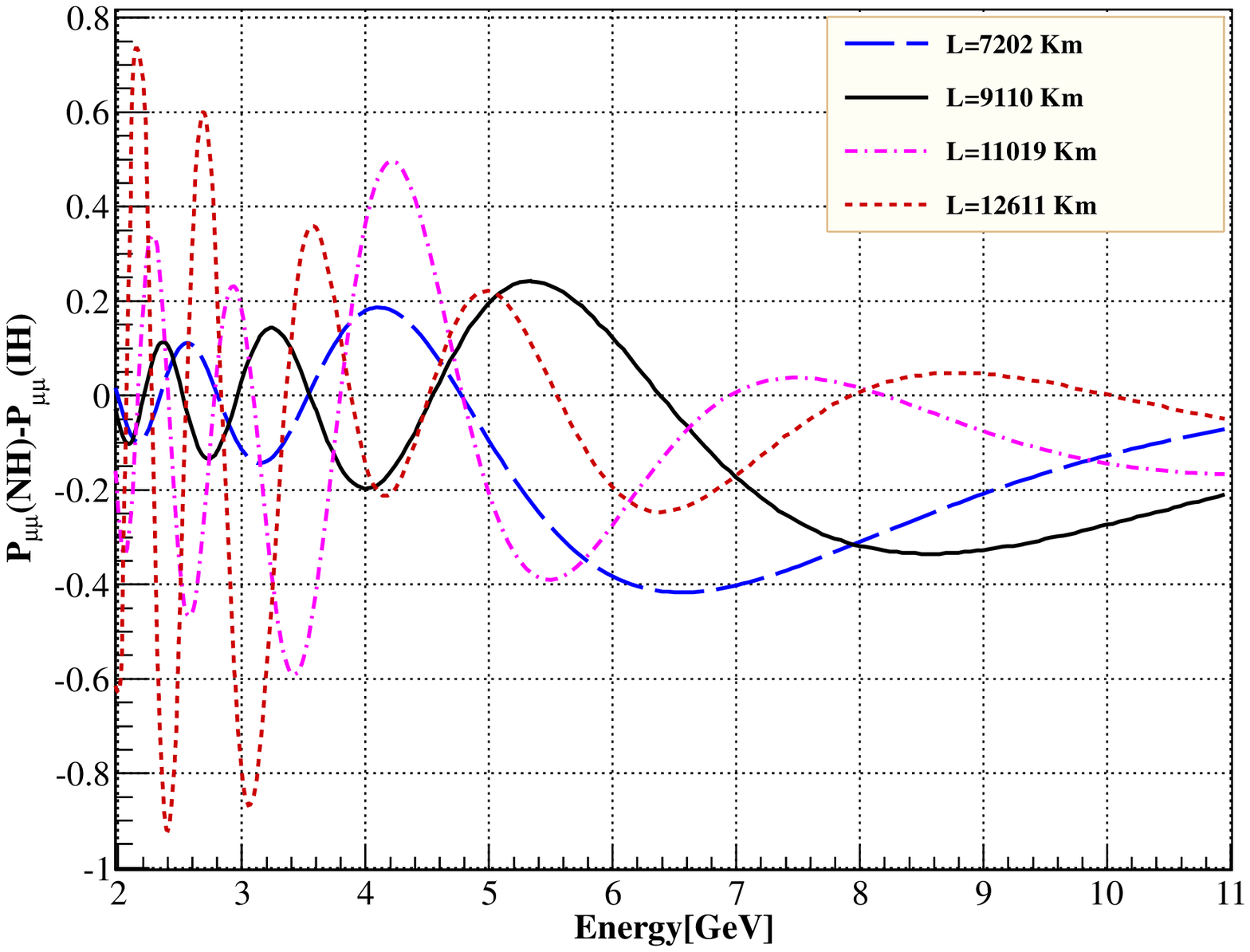}
\includegraphics[width=0.495\textwidth]{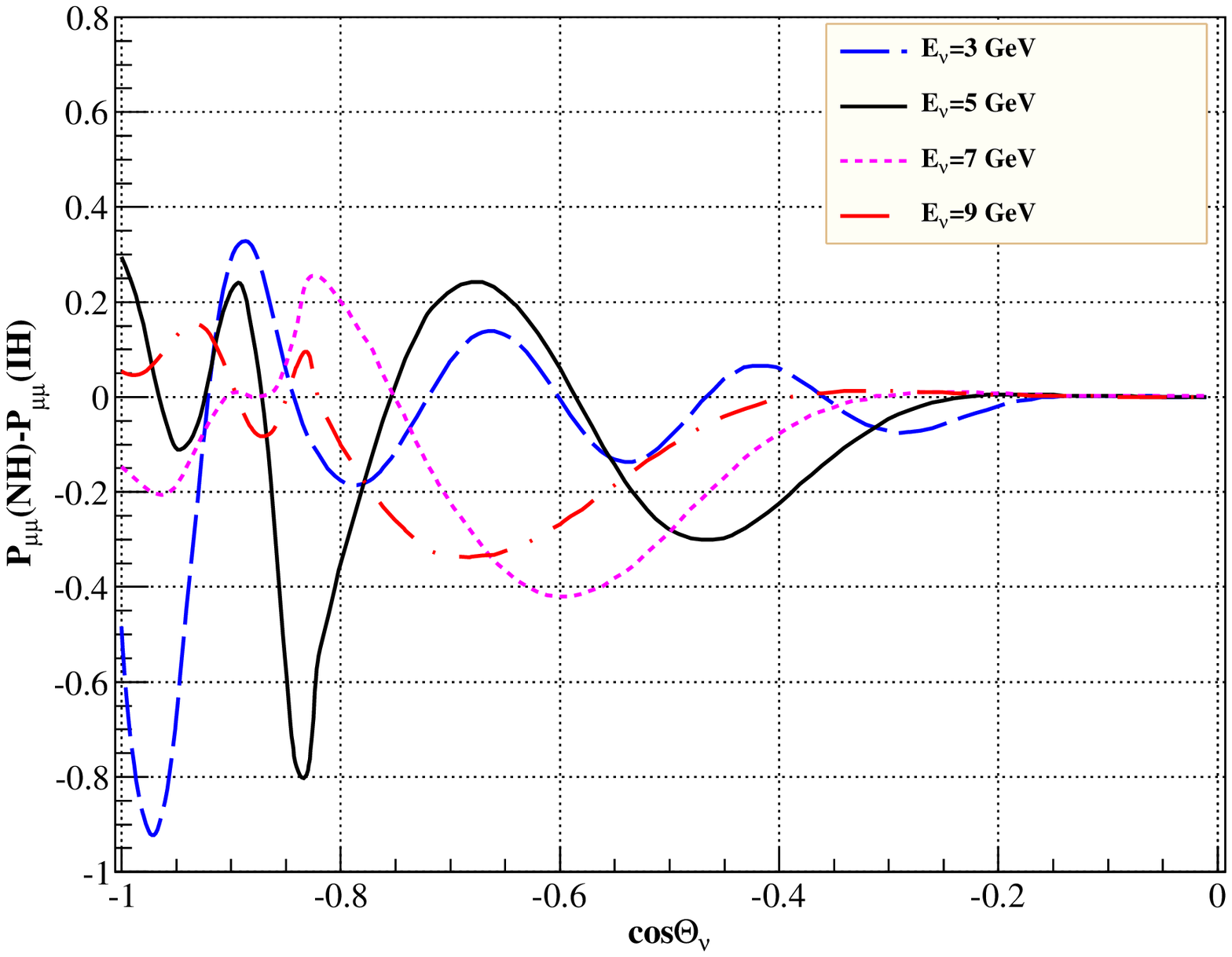}
\caption{The difference in the muon neutrino survival probability 
for normal and inverted hierarchy, $\Delta P_{\mu\mu}$, 
for various neutrino energy and zenith angles.
The left panel shows $\Delta P_{\mu\mu}$ as a function of the neutrino energy 
for four different choices of the zenith angle corresponding to four different path 
lengths for the neutrinos in the earth matter. The right 
panel shows $\Delta P_{\mu\mu}$ as a function of the neutrino zenith angle 
for four different choices of the neutrino energy. 
}
\label{fig:prob}
\end{figure}

\begin{table}[h]
\begin{center}
\begin{tabular}{|c|c|}
\hline
Parameter & True value used in data  \\[2mm] \hline
$\ms$ & $7.5 \times 10^{-5}$ eV$^2$ \\[2mm]
$\sin^2\theta_{12}$ & 0.3 \\[2mm]
$|\meff|$ & $2.4 \times 10^{-3}$ eV$^2$ \\[2mm]
$\delta_{CP}$ & 0 \\[2mm]
$\sa$ & 0.5  \\[2mm]
$\stch$ & 0.1    \\[2mm]
\hline
\end{tabular}
\caption{\label{tab:param}
Benchmark true values of oscillation parameters 
used in the simulations. 
}
\end{center}
\end{table}

We show in Fig. \ref{fig:prob} the 
difference between the survival probability predicted for the normal ($P_{\mu\mu}^{NH}$) and 
inverted ($P_{\mu\mu}^{IH}$) hierarchy. This figure has been generated 
by calculating the full three generation survival probability $P_{\mu\mu}$ 
at the benchmark oscillation 
parameter values given in Table \ref{tab:param} and for the 24 layer PREM 
profile of the earth \cite{prem}. 
We show the difference 
\be
\Delta P_{\mu\mu} = P_{\mu\mu}^{NH} - P_{\mu\mu}^{IH}
\,,
\label{eq:delpmm}
\ee
as a function of the neutrino energy in the left panel of Fig. \ref{fig:prob} for four  
different zenith angles of the neutrino -- 
the blue long-dashed line being for $L=7202$ km ($\cos\Theta_\nu=-0.56$),
the black solid line being for $L=9110$ km ($\cos\Theta_\nu=-0.71$), 
the pink dot-long-dashed line being for $L=11019$ km ($\cos\Theta_\nu=-0.86$), and 
the red short-dashed  line being for $L=12611$ km ($\cos\Theta_\nu=-0.99$). 
The right panel of Fig. \ref{fig:prob} shows this difference as a 
function of the neutrino zenith angle $\cos\Theta_\nu$, for four different 
neutrino energies -- the blue long-dashed line being for $E_\nu=3$ GeV, 
the black solid line being for $E_\nu=5$ GeV,
the pink short-dashed line being for $E_\nu=7$ GeV, and  
the red long-dashed line being for $E_\nu=9$ GeV. 
We notice that $\Delta P_{\mu\mu}$ fluctuates and changes sign as we change both 
the neutrino energy as well as the neutrino zenith angle. 
The sign of $\Delta P_{\mu\mu}$ is crucial, 
as it tells us whether earth matter effects increase or decrease the survival 
probability. 
Since $\Delta P_{\mu\mu}$ 
oscillates from positive to negative with energy and zenith angle, 
averaging it over either or both of these quantities will severely 
deplete and wash out the net earth matter effects in the $\numu$ signal 
at the detector. Therefore, observation of earth matter effects in $\numu$ 
is best performed with 
detectors having good energy as well as zenith angle resolution. 
\\

The fluctuations in $\Delta P_{\mu\mu}$ 
are seen to be faster for lower energies and longer baselines. 
In particular, for the zenith angles when the neutrinos cross the 
core of the earth, $\Delta P_{\mu\mu}$ is seen to be 
much more complicated. 
One can notice that for $E_\nu \sim 2-3$ GeV, 
earth matter effects are significantly stronger for these cases. 
This 
sharp increase in $|\Delta P_{\mu\mu}|$ is due to the so-called parametric 
enhancement of earth matter effects \cite{smirnov}, also known as 
oscillation length resonance effects \cite{petcov}.
While the matter effects are large here, the Fig. \ref{fig:prob} reveals that 
that the sign of $\Delta P_{\mu\mu}$ 
fluctuates very fast between being positive and negative, in this regime. 
This rapid fluctuation, as we will see later, will make the observation 
of earth matter effects extremely difficult at these energies and zenith angles. 
One would need extremely good reconstruction of the neutrino energy and angle 
for the observation of these parametric resonance effects. 
\\

Even for mantle crossing trajectories and/or higher 
neutrino energies, $\Delta P_{\mu\mu}$ 
fluctuates sign between positive and negative. However, 
the fluctuations are milder. 
Indeed from both the left as well as the right panel of Fig. \ref{fig:prob}
we note that for all baselines, $\Delta P_{\mu\mu}$ 
stays predominantly negative for neutrino energies $E_\nu \gtap 4-5$ GeV. 
In particular, we can see that $\Delta P_{\mu\mu} \simeq -0.4$ 
for a wide range of 
$E_\nu$ and $\cos\Theta_\nu$. In these cases the earth matter effects 
come mainly from the 
standard MSW  \cite{msw} enhancement of the oscillation probability. 
\\

The corresponding oscillation probabilities for the muon antineutrinos is same with the 
matter term $A$ replaced with $-A$ everywhere. Therefore, for the antineutrinos the 
effect of earth matter will be exactly opposite to what we see for the neutrinos. In particular, 
the antineutrino signal would see earth matter effects for the inverted hierarchy. 
If the antineutrino flux and cross-sections were same as the ones for the 
neutrinos, then adding up the data from neutrinos with those from the antineutrinos 
would completely wash out the earth matter effects between the normal and inverted 
hierarchy. However, in reality the antineutrino 
fluxes and cross-sections are smaller than those for the neutrinos, and hence 
a resultant mass hierarchy effect would survive even on adding the neutrino and antineutrino 
data. However, 
the magnetized iron calorimeters such as ICAL@INO, have excellent charge identification capabilities 
and hence can distinguish the neutrinos from the antineutrinos. Therefore, these detectors 
can observe earth matter effects separately in both the neutrino as well as the antineutrino channels.
Therefore, instead of partially washing the net earth matter effects, the antineutrino channel 
in these detectors add to the mass hierarchy sensitivity.

\section{Earth Matter Effects  in Event Rates}
\label{sec:ev}

We discussed the energy and zenith angle dependence of the earth matter effects in the muon neutrino 
survival probability $P_{\mu\mu}$ in the previous section. However, what arrives at the detector has 
impact from a 
combination of the survival probability $P_{\mu\mu}$ and the conversion probability $P_{e\mu}$. 
Though the survival probability $P_{\mu\mu}$ dominates, the conversion probability $P_{e\mu}$ 
has the effect of washing out partially the effect of earth matter and hence the mass hierarchy. 
In addition, the neutrinos interact with the detector nucleons through a charged current interaction to 
produce the corresponding muon and final state hadron(s). What is measured in the detector is the 
energy and zenith angle of the final state particles, {\it viz.}, the muon and, if possible, the hadron(s). 
A neutrino of energy $E_\nu$ could produce muons with any energy 
$E_\mu \leq E_\nu$. The same argument hold for the the zenith angle 
dependence as well. 
We had seen in the previous section that the net hierarchy effect in $P_{\mu\mu}$ 
oscillates between being positive and negative, both with energy as well as zenith angle. 
Therefore, in going from the neutrino energy to 
the muon energy through the interaction cross-section, what we effectively get is a 
smearing of the hierarchy dependent earth matter effects. 
This results in loss in the sensitivity of the experiment to the neutrino mass 
hierarchy. The finite detector resolutions for the energy and angle measurements further 
deteriorates the sensitivity. 
The only way to regain the sensitivity will be to simultaneously measure the energy 
and zenith angle of both the muon and the corresponding hadron, 
in order to reconstruct the neutrino energy and zenith angle. 
However, since the muon, and particularly the hadron, energy and 
angle cannot be 
measured very accurately in the iron calorimeter 
detector, the reconstruction ability of the experiment suffers. 
In this section we will show how the neutrino mass hierarchy sensitivity in the 
event spectrum changes with:
\begin{itemize}
\item going from measuring the events in terms of neutrino energy and zenith angle to 
muon and hadron energy and zenith angle,

\item inclusion of the appearance channel $P_{e\mu}$,

\item inclusion of the finite detector resolutions.

\end{itemize}
We show the dependence of mass hierarchy sensitivity on these 
issues first in the muon event spectrum, 
then in the hadron event spectrum and finally in the neutrino event spectrum.

\subsection{Earth Matter effects in Muon Events}
\label{sec:evmuon}

\begin{figure}[p]
\centering
\includegraphics[width=0.495\textwidth]{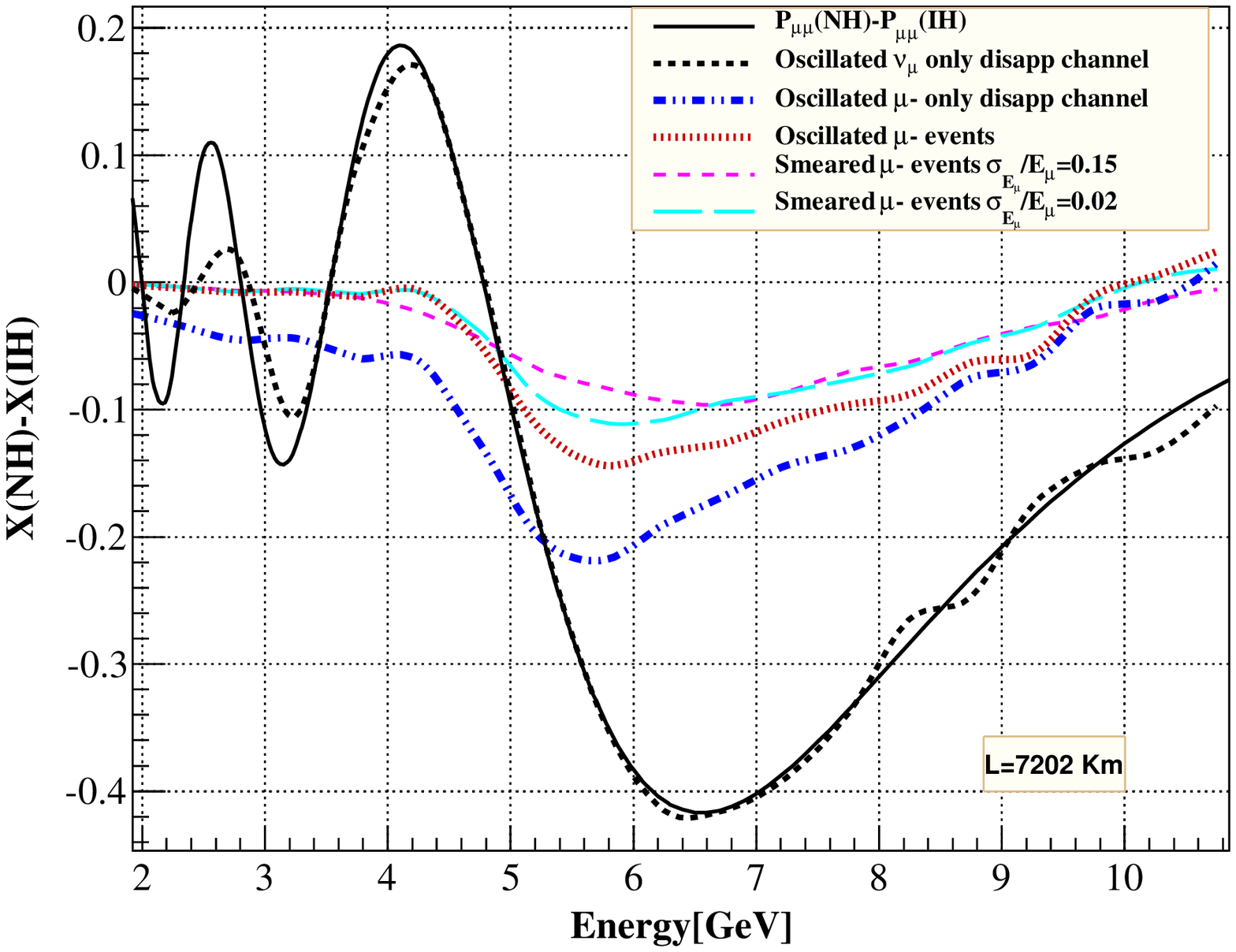}
\includegraphics[width=0.495\textwidth]{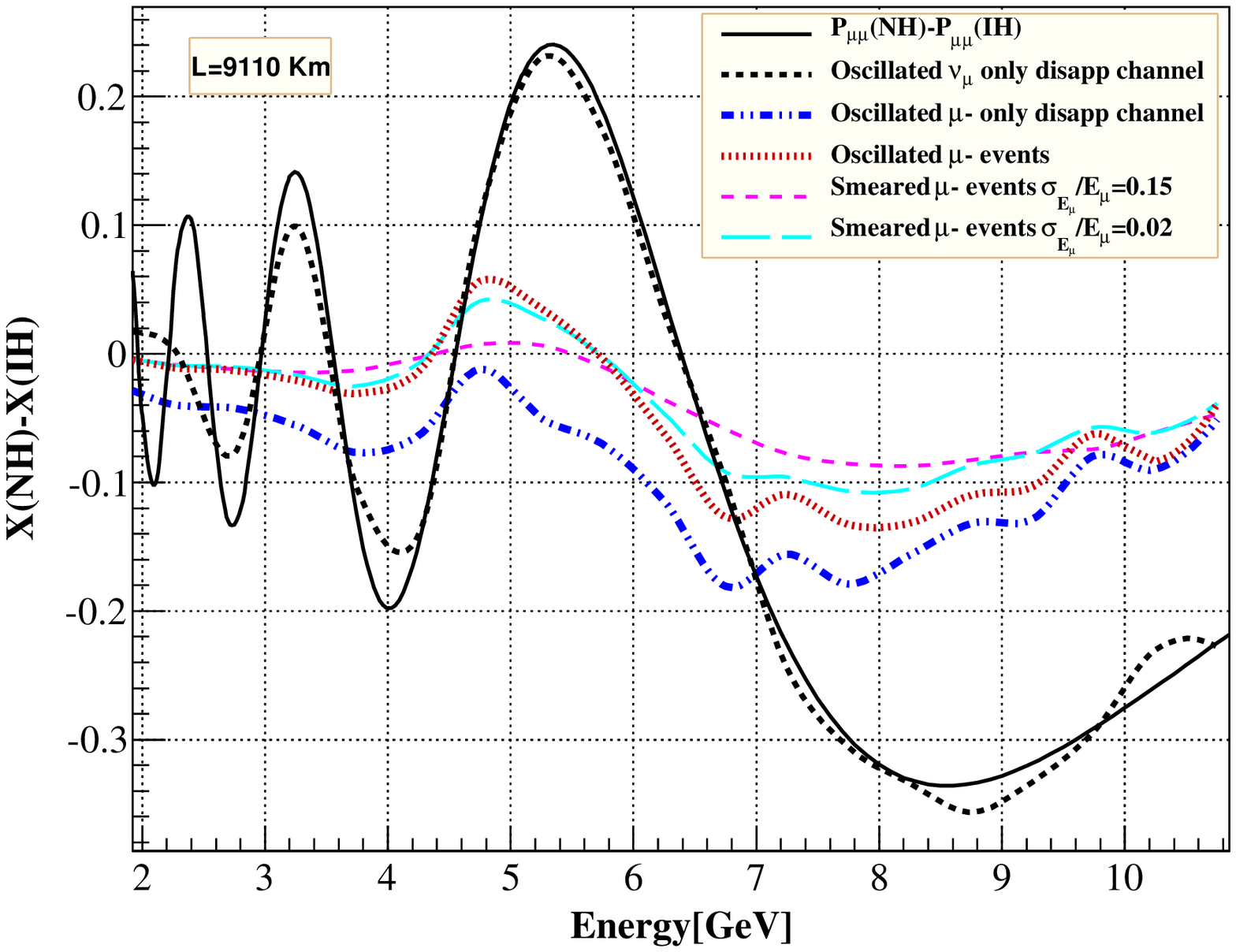}

\includegraphics[width=0.495\textwidth]{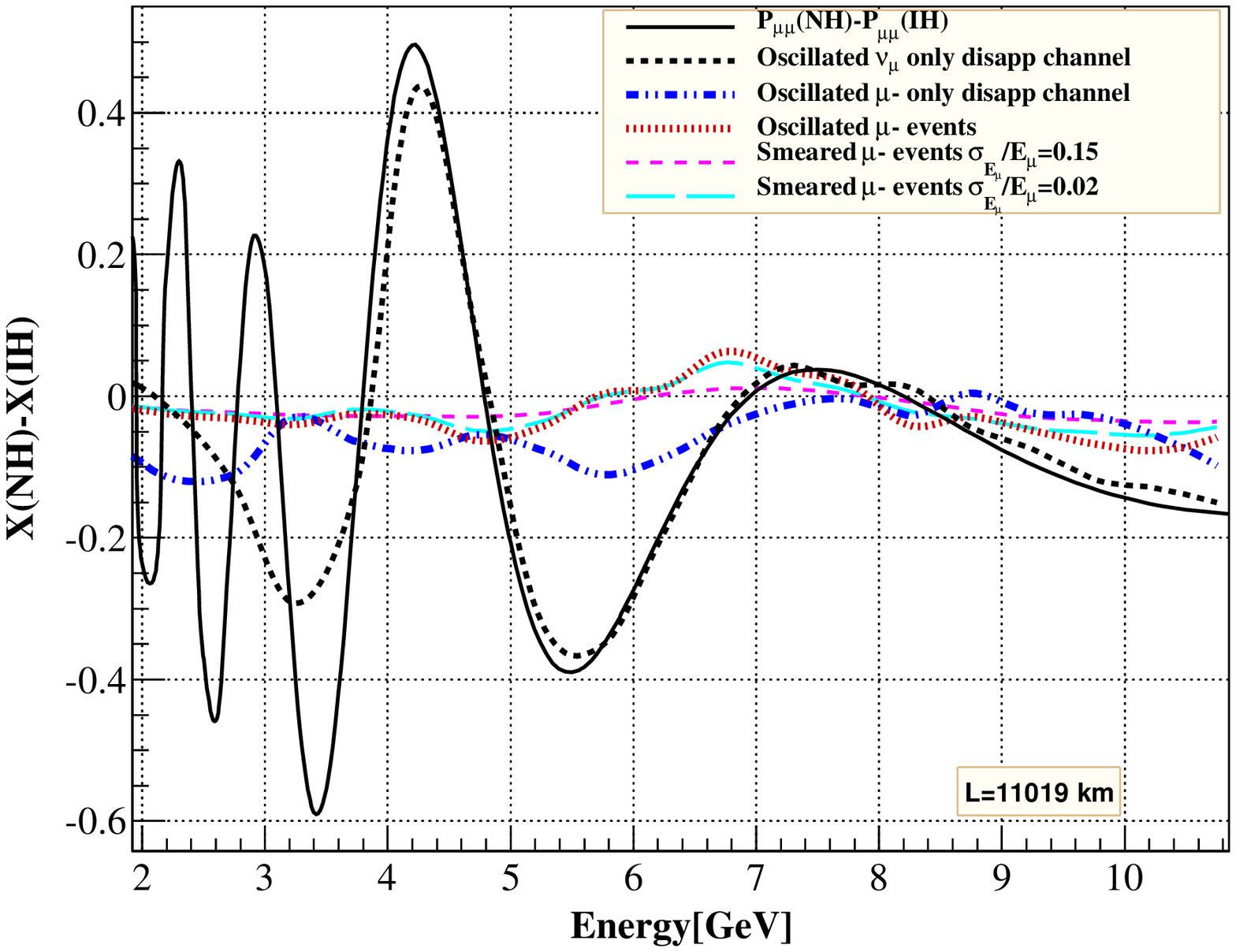}
\includegraphics[width=0.495\textwidth]{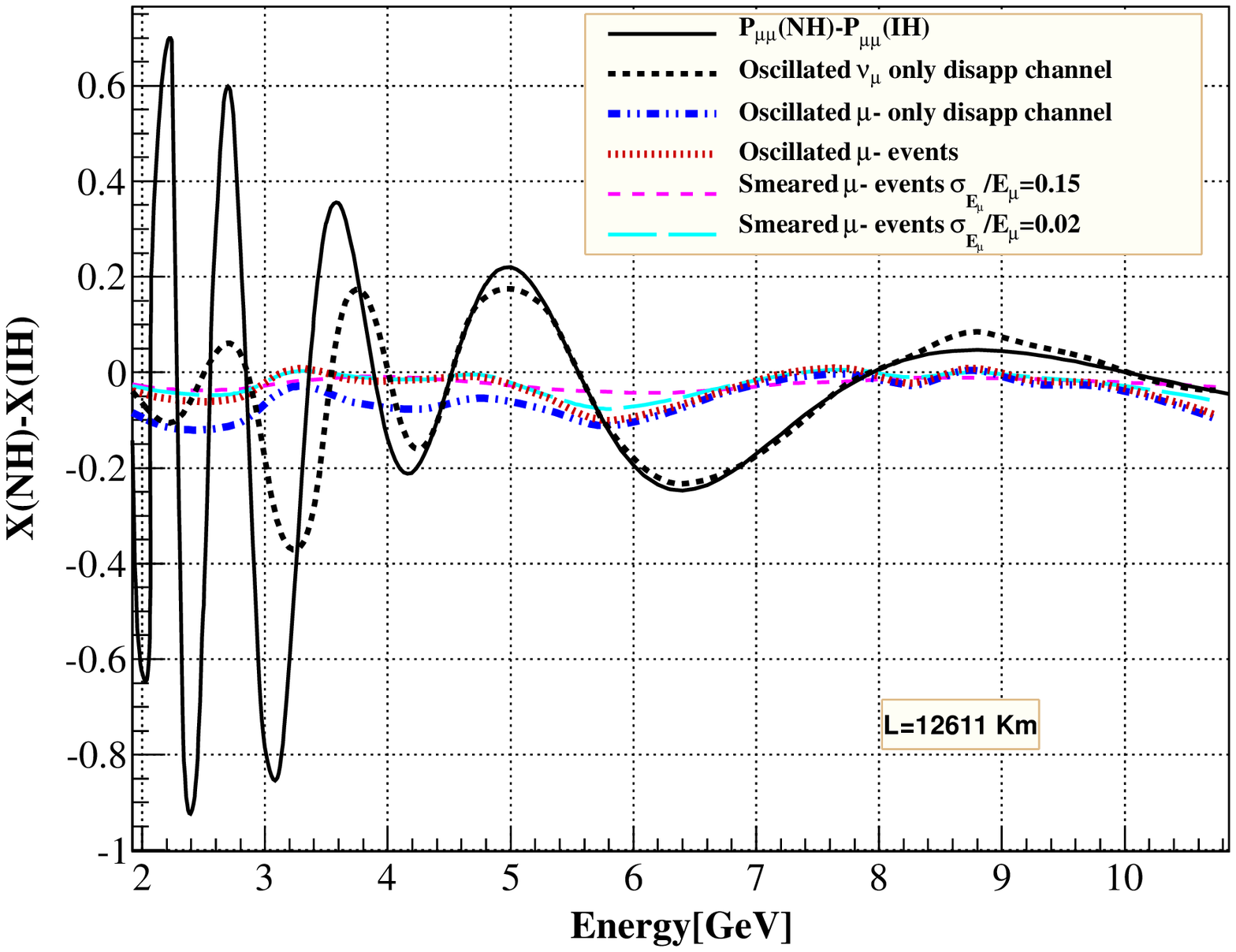}
\caption{The difference in the predicted muon event rates for normal and 
inverted hierarchy, normalized to the no-oscillation muon event rate, shown as 
a function of energy. The four panels show the 
event spectrum generated in four zenith angle 
bins, marked by the path-length traversed at the mid-point of the bin. 
The solid black lines show the $\Delta P_{\mu\mu}$ as in Fig. \ref{fig:prob}. The 
black dashed lines give the difference between the predicted rate for normal and inverted 
hierarchy when only the $P_{\mu\mu}$ (disappearance) 
channel is taken and data binned in neutrino energy.
The blue dot-dashed lines show the corresponding difference when only the 
$P_{\mu\mu}$ channel is taken and data binned in muon energy. The red dotted lines are 
obtained when we add the $P_{e\mu}$ (appearance) channel and bin the data in muon 
energy. The pink dashed and cyan long dashed lines are obtained when we 
apply on the red dotted lines 
the muon energy resolution functions with widths $\sigma_{E_\mu}/E_\mu = 0.15\%$ and 
2\%, respectively. 
}
\label{fig:evmuon}
\end{figure}

For generating the unoscillated $\mu^-$ (and $\mu^+$) event spectrum, we have used the 
NUANCE event generator \cite{nuance} with Honda et al. fluxes \cite{honda}. 
We use a re-weighting 
algorithm on these raw events to obtain the oscillated event spectrum.
This is then folded with the detector 
response functions in order to simulate the measured event spectrum in the 
detector as follows:
\footnote{Details of how we simulate the atmospheric neutrino events in the detector can be found in  
\cite{mh,precision}. }
\be
N_i^{\prime th} (\mu^-) =  {\cal N}
\sum_i
\int_{E_\mu^{min}}^{E_\mu^{max}}  \!\!\!dE_\mu^\prime
\int_{\cos\Theta_\mu^{min}}^{\cos\Theta_\mu^{max}}  \!\!\!d\cos\Theta_\mu^\prime
\,\, \,
R_E\,
R_{\Theta}\,
\bigg({\cal E}_i{\cal C}_i\,n_i(\mu^-) + {\overline {\cal E}_i}(1-\overline{{\cal C}_i})\,n_i(\mu^+)\bigg)\,
\,,
\label{eq:eventsth}
\ee
where $n_i(\mu^-)$ and $n_i(\mu^+)$ 
are the number of oscillated $\mu^-$ and $\mu^+$ events  respectively,
in the $i^{th}$ bin, and 
${\cal N}$ corresponds to a specific exposure for the experiment. 
An expression similar to Eq. (\ref{eq:eventsth}) 
can we written for the $\mu^+$ events $N_i^{\prime th} (\mu^+)$. 
The summation is over $i$ where $i$ scans all true 
muon energy and muon zenith angle bins. The distinction between the 
true and measured parameters are as follows. 
The quantities $E_\mu$ and $\cos\Theta_\mu$ are the 
true (kinetic) energy and true zenith angle of the muon, while 
$E_\mu^\prime$ and $\cos\Theta_\mu^\prime$ are the 
corresponding measured (kinetic) energy and measured zenith angle obtained from the 
observation of the muon track in the detector. 
The reconstruction efficiencies of $\mu^-$ and $\mu^+$  in the $i^{th}$ bin 
are given by ${\cal E}_i$ and $\overline{{\cal E}_i}$ respectively, 
while ${\cal C}_i$ and $\overline{{\cal C}_i}$ are the 
corresponding charge identification efficiencies. 
We take the resolution functions 
$R_E$ and $R_{\Theta}$ 
to be Gaussian, 
\be
R_E = \frac{1}{\sqrt{2\pi}\sigma_{E_\mu}} \exp \bigg(
\frac{-(E_\mu^\prime - E_\mu)^2}{2\sigma^2_{E_\mu}}\bigg )
\,,
\label{eq:eresoln}
\ee
\be
R_{\Theta} =  \frac{1}{\sqrt{2\pi}\sigma_{\Theta_\mu}}\exp \bigg(
\frac{-(\cos\Theta_\mu^\prime - \cos\Theta_\mu)^2}{2\sigma^2_{\Theta_\mu}}\bigg )
\,,
\label{eq:zenithresoln}
\ee
respectively. While the reconstruction efficiencies, charge identification 
efficiencies as well as the resolution functions could be 
function of both the muon energy and muon zenith angle \cite{inomuon,mh,precision}, 
in this paper we will use flat projected values for these parameters. 
This is justified as the results presented here are for the sake of illustration only 
to show the dependence of the neutrino mass hierarchy to the 
different ways of treating the data and their corresponding 
detector response functions. Since the first set of simulation results 
performed by the INO collaboration have shown extremely good 
muon zenith angle resolution, we will fix $\sigma_{\Theta_\mu} = 0.01$ for 
all muon energies and zenith angles, throughout this paper. The charge identification efficiency 
is also fixed at flat 99\% everywhere. The muon energy resolution and the 
reconstruction efficiencies are allowed to vary. We will mention this as and when 
applicable. All detector response functions are taken to be the same for the 
$\mu^-$ and $\mu^+$ events. 
\\

As discussed before, for normal (inverted) hierarchy we expect earth matter 
effects in the $\mu^-$ ($\mu^+$) events, while for inverted (normal) hierarchy 
there will be no earth matter effects in the $\mu^-$ ($\mu^+$) channel. 
To quantify the mass hierarchy sensitivity of the experiment, we simulate the 
$\mu^-$ and $\mu^+$ data separately for the normal mass hierarchy
and fit this data with the inverted mass hierarchy. 
Therefore for this example case, there will be earth matter effects in the $\mu^-$ 
events in the data, while no earth matter effects in the theoretical fit. On the other hand, 
for the $\mu^+$ channel, there will be no earth matter effects in the data sample, but 
the theoretical prediction for the event spectrum will have earth matter effects. 
Because the magnetized iron calorimeter such as ICAL@INO will 
have excellent charge identification capabilities, it can distinguish the $\mu^-$ signal 
from the $\mu^+$ one.
What is relevant for mass hierarchy studies will then be the difference between the 
predicted events for normal and inverted mass hierarchies, separately in the 
$\mu^-$ and $\mu^+$ channels. 
\\

We present this difference for $\mu^-$ events 
in Fig. \ref{fig:evmuon}. A similar figure can be drawn for the $\mu^+$ events, but 
we will not repeat it here. 
The four panels of this figure are for 
four different zenith angles, and hence, four different path lengths of the neutrino in earth, 
$L=7202$ km, 9110 km, 11019 km and 12611 km. 
The different lines in this figure are 
as follows. The black solid lines give $\Delta P_{\mu\mu} = P_{\mu\mu}^{NH} - P_{\mu\mu}^{IH}$, 
(cf. Eq. \ref{eq:delpmm}) as a function of the {\it neutrino energy} $E_\nu$.  
These are same as those shown in the left panel of Fig. \ref{fig:prob} and have been 
repeated here to show the impact of the interaction cross-section on 
the mass hierarchy sensitivity. 
All other lines in this figure show the difference between the predicted 
events for normal and inverted hierarchies, normalized to the unoscillated events.  
Each muon event output from Nuance is characterized in terms of the incoming 
true neutrino angle and energy and outgoing true muon angle and energy. 
Therefore, we can bin the data in either the neutrino energy or the muon 
energy. 
The  black dashed lines show this difference when the data is binned in 
true neutrino energy and angle bins and only the muon neutrino survival probability $P_{\mu\mu}$ 
is taken into account. The x-axis for these lines are therefore the 
true neutrino energy $E_\nu$. 
We can see 
that the the black dashed lines follow the black solid lines to reasonable accuracy. 
This is expected since we have used only $P_{\mu\mu}$ for the event difference shown by 
the  black dashed lines, and since it is binned in terms of the neutrino energy and angle, 
it has a one-to-one correspondence with the difference $\Delta P_{\mu\mu}$ shown by the 
back solid lines. The small difference between the two sets of curves comes mainly due to the 
Monte Carlo fluctuations in the NUANCE output. 
\\

Next we use the same data set where only the 
survival probability $P_{\mu\mu}$ is considered and bin the data in terms of the 
{\it true muon energy $E_\mu$}. 
This is shown by the blue dot-dashed curves in Fig. \ref{fig:evmuon}. 
The x-axis for these lines are therefore the 
true muon energy $E_\mu$. 
Notice now 
that the event spectrum now gets degraded in energy. For example for $L=7202$ km case, 
while the maximum difference due to earth matter effects were coming at $E_\nu \sim 6-7$ GeV in terms of the 
true neutrino energy, it appears at $E\sim 5-6$ GeV in terms of the true muon energy. 
Notice also that the net earth matter effect is also substantially reduced in going from neutrino 
energy to muon energy. This mainly comes due to the kinematic averaging effect of the
neutrino-nucleon cross-sections wherein a neutrino with a given neutrino energy $E_\nu$ 
could produce a muon with any energy between $E_\nu$ and zero. This results in 
the smearing out of the oscillation effects of $\Delta P_{\mu\mu}$ and since the 
$\Delta P_{\mu\mu}$ fluctuates between being positive and negative, the averaging 
brings about a net cancellation of the earth matter effects. 
However, despite this kinematic smearing of the signal, some residual earth matter effects 
remain with $\Delta X = X(NH)-X(IH) <0$ predominantly. 
In particular, for the mantle-crossing bins shown in upper panels for $L=7202$ km and 9110 km in Fig. 
\ref{fig:evmuon}, the $\Delta X <0$ for all muon energies and 
we get $\Delta X  \simeq -0.2$ at $E_\mu \simeq 5-6$ GeV and $6-8$ GeV, 
respectively. Since the oscillations in the net earth matter effects are 
larger for the core-crossing bins shown in the lower panels for $L=11019$ km and 12611 km in Fig. 
\ref{fig:evmuon}, the smearing is more pronounced for these zenith angles. 
As a result, even though the actual earth matter effects in the neutrinos are much larger for these 
zenith angles, in the muon sample the averaged $\Delta X $ for these bins are significantly lower.
Still there is some residual earth matter effects with  $\Delta X  \simeq -0.1$ predominantly, for 
most muon energies. 
\\

The blue dot-dashed lines that we discussed above have contribution from only the 
survival probability channel $P_{\mu\mu}$. However, since 
both $\numu$ and $\nue$ are produced in the earth's atmosphere, some of the 
$\numu$ arriving at the detector will be the ones produced as $\nue$ and which 
have oscillated into $\numu$ through the $P_{e\mu}$ conversion probability. 
These ``appearance events" have to be added to the ones we had obtained 
using the survival probability $P_{\mu\mu}$, to obtain the final muon event spectrum which 
will be observed in magnetized iron detectors such as ICAL@INO. 
These are shown  by the red dotted lines in Fig. \ref{fig:evmuon}.
A comparison of the red dotted lines with the blue dot-dashed lines reveals the 
impact of the $P_{e\mu}$ channel on the net earth matter effects and hence 
the mass hierarchy sensitivity of atmospheric neutrino experiments. The 
effect of $P_{e\mu}$ is to reduce $|\Delta X|$ for all energies and all 
zenith angles. The reason for this can be seen from 
comparing the simplified expressions given 
in Eqs. (\ref{eq:pmm}) and (\ref{eq:pem}). We can see that one can write 
\be
P_{\mu\mu}^{approx} =  1 \!\!\!\!\!\!\!\!\!\!&&
-\sin^2\theta_{13}^M\sin^22\theta_{23} \sin^2\frac{[(\ma + A)-(\ma)^M]L}{8E_\nu}
\\ \nonumber
&& - \cos^2\theta_{13}^M\sin^22\theta_{23} \sin^2\frac{[(\ma + A)+(\ma)^M]L}{8E_\nu}
\\
&&-\sin^2\theta_{23} P_{e\mu}^{approx}
\,.
\label{eq:pmm1}
\ee
We note that $P_{e\mu}$  appears with a negative 
sign in the approximate expression for $P_{\mu\mu}$. 
Since the resultant muon flux at the detector is a {\it sum} of the 
$P_{\mu\mu}$ and $P_{e\mu}$ probabilities multiplied by the 
corresponding atmospheric neutrino fluxes, 
the contribution to earth matter effects coming from $P_{\mu\mu}$ 
get partially cancelled 
with that coming from $P_{e\mu}$. In Fig. \ref{fig:evmuon} 
this is reflected in the reduction of $|\Delta X|$ 
when going from the toy case where only $P_{\mu\mu}$ driven events were considered 
(blue dot-dashed lines) to the realistic case where both 
$P_{\mu\mu}$  and $P_{e\mu}$ are taken (red dotted lines) into account. 
In particular, for the $L=7202$ km case, the $\Delta X$ changes from 
$-0.2$ to $-0.1$ at $E_\mu \simeq 5-6$ GeV. 
From the figure one can see the reduction in mass hierarchy 
sensitivity for other baselines as well. 
\\

Finally, we impose the muon energy resolution on the event sample. 
This, as expected, brings about a further smearing of the energy spectrum. 
We use a Gaussian energy resolution function for the muons as given by 
Eq. (\ref{eq:eresoln}). 
We show the impact of this smearing in Fig. \ref{fig:evmuon} by the cyan long-dashed lines 
($\sigma_{E_\mu} = 0.02E_\mu$) and the pink short-dashed lines 
($\sigma_{E_\mu} = 0.15E_\mu$).
The effect of putting the muon energy resolution function is obviously to further 
smear the energy spectrum and the higher the $\sigma_{E_\mu}/E_\mu$, the higher will be the smearing, 
as seen in the figure. Nevertheless, we can see that even after imposing the 
energy resolution we have a residual $\Delta X$ which can be used 
to distinguish the normal from the inverted hierarchy. It is this final residual 
$\Delta X$ which translates into the $\Delta \chi^2$ for the wrong neutrino mass 
hierarchy in our statistical analysis. 

\subsection{Earth Matter Effects in Hadron Events}
\label{sec:evhadron}

The charge current interaction of $\numu$  in the detector produces a hadron (or a bunch of 
hadrons) in addition to the muon. While the muon moves over long distances making 
long tracks in the detector, the hadron(s) produce(s) a shower. 
The INO collaboration has 
performed their first set of simulations studying the response of the ICAL detector to 
hadrons. The results showing the calibration of the detector to hadron energy and 
the corresponding hadron energy resolution have been presented in \cite{inohadron}.
The study of the hadron response of the MONOLITH detector proposal was made in detail, 
both in computer simulations as well as by putting the prototype in a test beam 
\cite{monolithenergy,monolithangle}. The energy and angle resolution of 
the hadron shower has also been studied extensively by the MINOS collaboration. 
In all works so far, the information from the hadrons have been used to reconstruct the 
neutrino energy and angle. However, we propose a different method of treating the 
hadron data. We will take the hadron data at par with the muon data and add their contribution 
to the statistical analysis of the mass hierarchy sensitivity of the magnetized iron detector. 
To the best of our knowledge, this has not been done before. 
\\

In order to add the hadron contribution to the mass hierarchy sensitivity, we first need to 
bin the hadron data in a suitable way. 
In this paper we treat the hadron event sample of magnetized iron detectors as follows. Since every 
hadron shower is associated with a corresponding muon coming from the same charged 
current interaction vertex of the neutrino, we use the muon as a tag for the hadron event. 
Since the muon zenith angle is reconstructed extremely well, we tag the hadron 
with the zenith angle of the corresponding muon. 
This means that for every muon in a given muon zenith angle bin, 
we group together all the corresponding hadrons. We next bin this group in terms of 
the hadron energy.  
Note that the most obviously way to calculate the true energy in the hadrons would be by adding 
up the energy of the hadrons in the final state for each event. 
When handling real data, this is what will be done. 
However, in our analysis we do not have real data. We only use the output of the event generator 
as our simulated data. Here a complication arises due to the fact that 
the final state provided by NUANCE has a large number 
of additional 
hadrons, which come from the breaking of the iron nucleus. Hence, adding them 
all up does not directly help in finding the energy released in the hadron due to the charged 
current interaction. 
Therefore instead of doing that, we calculate $E_H = E_\nu - E_\mu$, where $E_\nu$ 
is the true energy of the incoming neutrino and $E_\mu$ is the true energy of the 
muon, provided by the event generator. We call $E_H$ the true energy in the hadrons. 
Having calculated the energy of all the associated hadrons in a given muon zenith angle bin, 
we redistribute these events in hadron energy bins, starting from 0.5 GeV to 
10.5 GeV. 
\\

Following this methodology for hadron binning, 
the number of hadron events in each hadron energy bin can be written as  
\be
H_i^{\prime th} (\mu^-) =  {\cal N}
\sum_i
\int_{E_H^{min}}^{E_H^{max}}  \!\!\!dE_H^\prime
\int_{\cos\Theta_\mu^{min}}^{\cos\Theta_\mu^{max}}  \!\!\!d\cos\Theta_\mu^\prime
\,\, \,
R_{E_H}\,
R_{\Theta}\,
{\cal E}^H_i\bigg({\cal E}_i{\cal C}_i\,h_i(\mu^-) + \overline{{\cal E}}_i(1-\overline{{\cal C}_i})\,h_i(\mu^+)\bigg)\,
\,.
\label{eq:evhadron}
\ee
Since we use the muon events as a tag and calculate the hadron energy binned 
data for every muon zenith angle bin, we calculate the hadron events corresponding to the 
$\mu^-$ events (denoted as $h_i(\mu^-)$ in Eq. (\ref{eq:evhadron})) as well as the 
ones associated with the $\mu^+$ mis-identified as $\mu^-$ 
(denoted as $h_i(\mu^+)$ in Eq. (\ref{eq:evhadron})). The quantities 
${\cal C}_i$ and $\overline{{\cal C}_i}$ are the charge identification efficiencies of 
$\mu^-$ and $\mu^+$ events, respectively, as defined in Eq. (\ref{eq:eventsth}). 
The reconstruction efficiency of the hadron shower is given by ${\cal E}^H$, and 
in this paper it is taken to be 
the same for hadrons associated with both $\mu^-$ and $\mu^+$. In the way we are handling 
the hadron events, we also multiply our events with ${\cal E}_i$ and $ \overline{{\cal E}}_i$ 
for the $\mu^-$ and $\mu^+$ events respectively, since we first look at events which 
already have a muon track reconstructed. From this set of events we find the subset for 
which even the hadron shower can be reconstructed.  
$R_E^H$ is the energy resolution function of the hadrons, for which we use a Gaussian 
function similar to Eq. (\ref{eq:eresoln}) 
\be
R_{E_H} = \frac{1}{\sqrt{2\pi}\sigma_{E_H}} \exp \bigg(
\frac{-(E_H^\prime - E_H)^2}{2\sigma^2_{E_H}}\bigg )
\,,
\label{eq:eresolnhad}
\ee
where 
$E_H$ and 
$E_H^\prime$ correspond to the true energy and measured energy of the hadrons 
respectively. We need to put $R_\Theta$ which is the muon zenith 
angle resolution given by Eq. (\ref{eq:zenithresoln}) in the above since 
the hadrons are binned inside the muon zenith angle bin. Note that the 
Eq. (\ref{eq:evhadron}) gives the data binned in hadron energy only. 
When we bin the data in both the hadron energy as well as zenith angle, 
we will have to introduce the integral for the measured 
hadron angle bin and the corresponding hadron angle resolution function. We will 
discuss this later in section \ref{sec:chihad2}.
\\

\begin{figure}[p]
\centering
\includegraphics[width=0.495\textwidth]{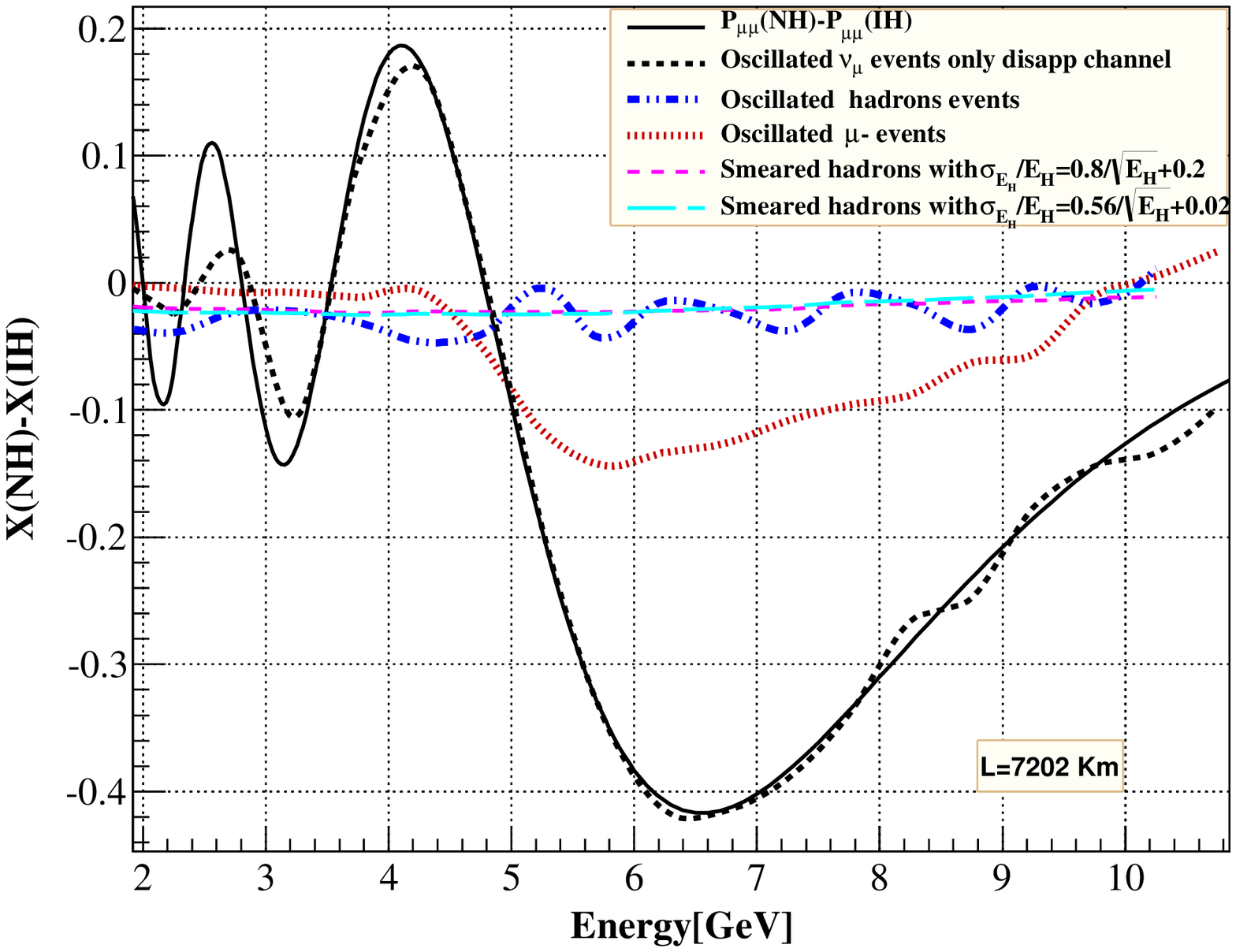}
\includegraphics[width=0.495\textwidth]{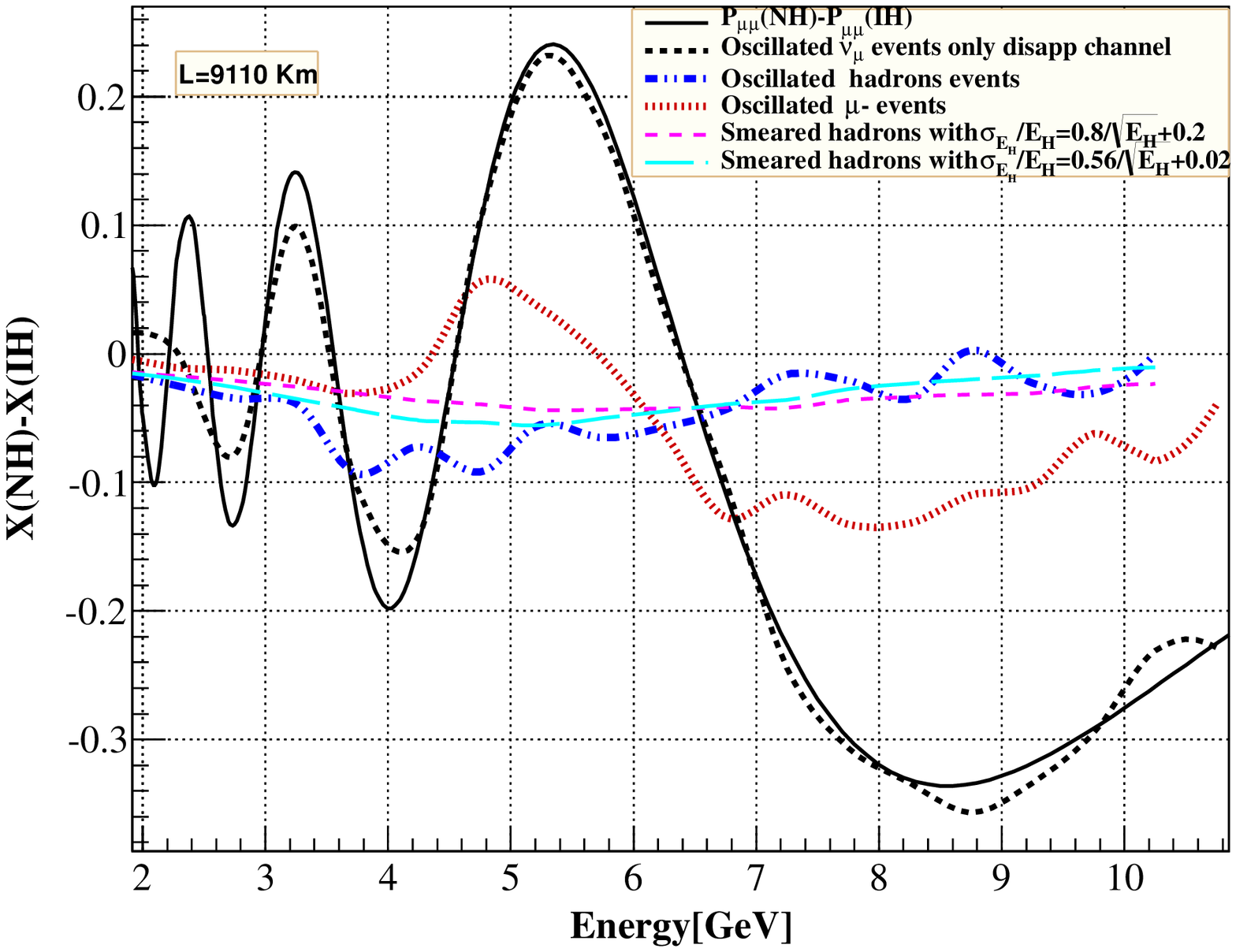}

\includegraphics[width=0.495\textwidth]{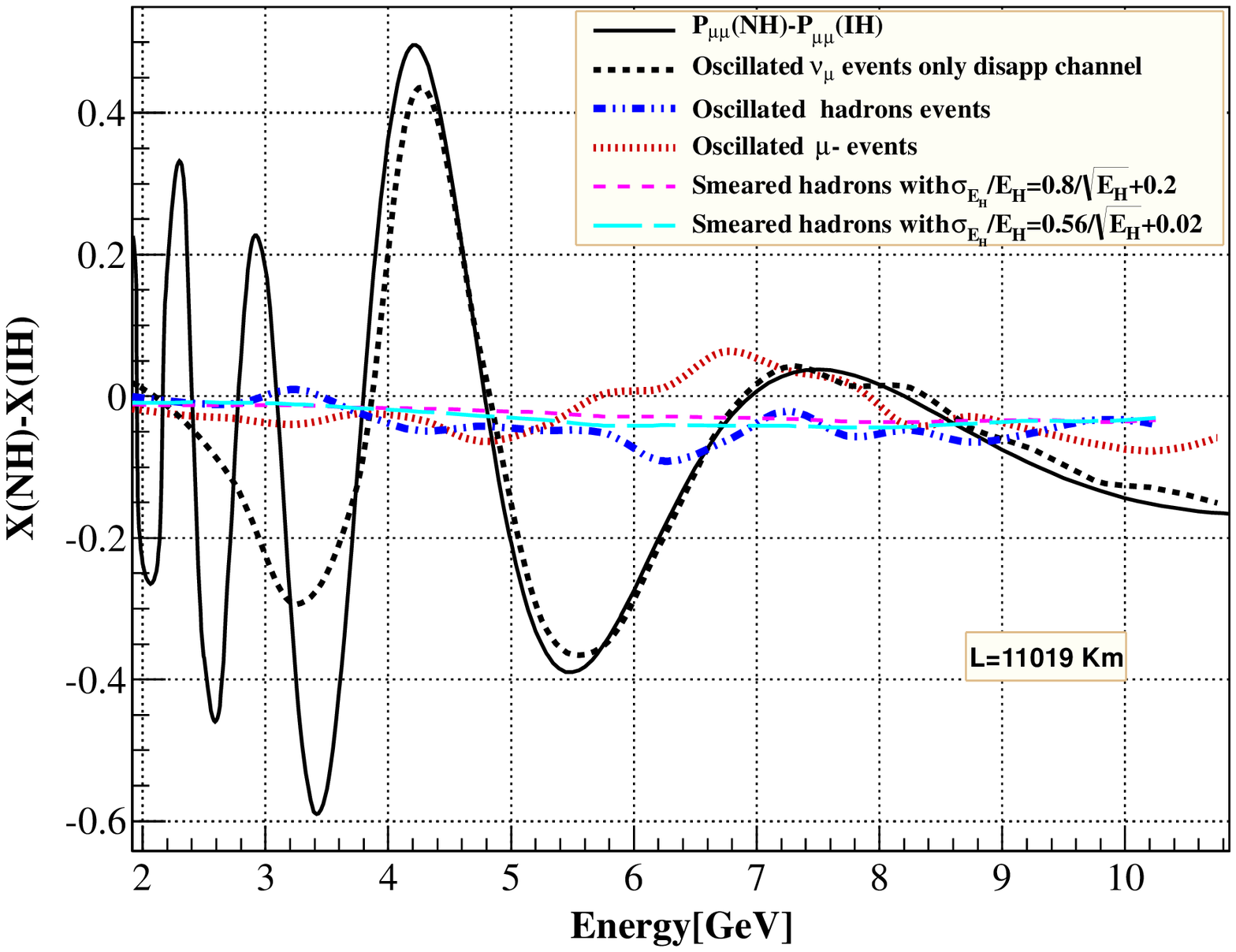}
\includegraphics[width=0.495\textwidth]{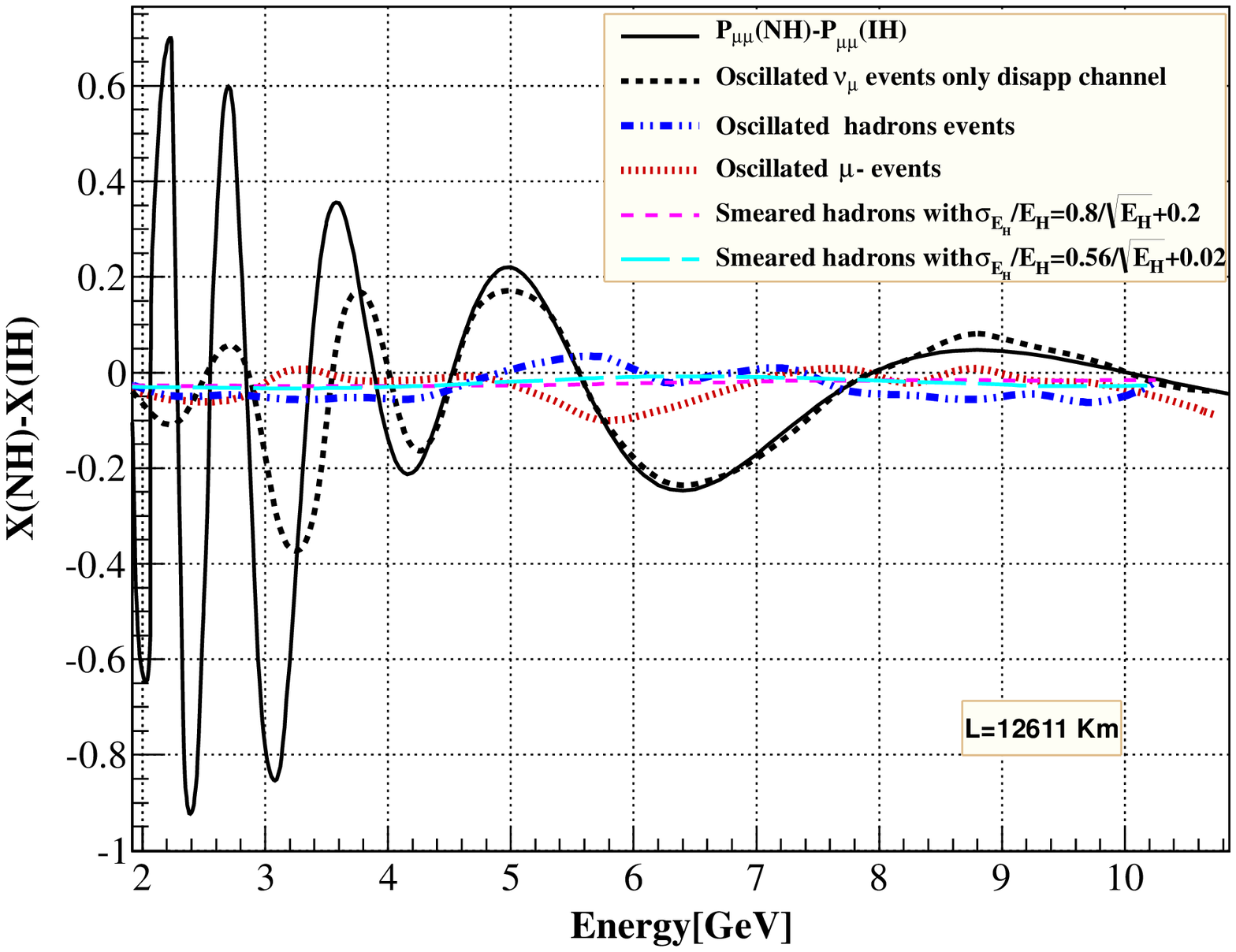}
\caption{The difference in the predicted hadron event rates for normal and 
inverted hierarchy, normalized to the no-oscillation hadron event rate, shown as 
a function of energy. The four panels show the 
event spectrum generated in four zenith angle 
bins, marked by the path-length traversed at the mid-point of the bin. 
The solid black lines, black dashed lines and red dotted lines 
show the things as in Fig. \ref{fig:evmuon}. The 
black dashed lines give the difference between the predicted rate for normal and inverted 
hierarchy when only the $P_{\mu\mu}$ (disappearance) 
channel is taken and data binned in neutrino energy.
The blue dot-dashed lines show the corresponding difference when only the 
$P_{\mu\mu}$ channel is taken and data binned in muon energy. The red dotted lines are 
obtained when we add the $P_{e\mu}$ (appearance) channel and bin the data in muon 
energy. The pink dashed and cyan long dashed lines are obtained when we 
apply on the red dotted lines 
the muon energy resolution functions with widths $\sigma_{E_\mu}/E_\mu = 0.15\%$ and 
2\%, respectively. 
}
\label{fig:evhadron}
\end{figure}

In Fig. \ref{fig:evhadron} 
we show this binned data for the hadrons for four muon zenith angle bins corresponding 
to path lengths of $L=7202$ km (top left panel), $L=9110$ km 
(top right panel), $L=11019$ km (bottom left panel), 
and $L=12611$ km (bottom right panel). 
The hadron energy spectrum, 
without putting any energy resolution is 
shown by the blue dot-dashed lines in Fig. \ref{fig:evhadron}. 
We can see that even for the hadrons $\Delta X<0$ predominantly, 
for all hadron energies. In fact, we can see that $\Delta X \simeq -0.05$ for a 
wide range of hadron energies, for all muon zenith angle bins. The pink short-dashed and the 
cyan long-dashed lines show the hadron spectrum smeared by $\sigma_{E_H}/E_H = 0.8/\sqrt{E_H}+0.2$ 
and $\sigma_{E_H}/E_H = 0.56/\sqrt{E_H}+0.02$, respectively. Since the shape of the 
hadron energy spectrum shown by the blue dot-dashed lines was anyway almost flat, the 
further smearing of spectrum due to the finite energy resolution of the detector 
does not bring about any drastic change in the spectral shape. Indeed the 
spectra for $\sigma_{E_H}/E_H = 0.8/\sqrt{E_H}+0.2$ 
and $\sigma_{E_H}/E_H = 0.56/\sqrt{E_H}+0.02$ look almost identical to each other. 
As in Fig. \ref{fig:evmuon}, the black solid lines 
in all the panels show the $\Delta P_{\mu\mu}$ at the probability level. 
Likewise, the  black dashed lines correspond to difference in the 
event spectrum between normal and inverted hierarchies in terms of the 
neutrino energy with only the disappearance channel. The red dotted lines in all the panels show the 
event spectrum in terms of the true muon energy, where we have included events from 
both the disappearance as well as appearance channels but have not incorporated the 
muon energy resolutions. These lines again are the same as the red dotted lines in 
Fig. \ref{fig:evmuon}. Therefore, the black solid lines, black dashed lines and the 
red dotted lines and have been shown here again for comparison of the hadron spectrum 
with the corresponding neutrino and the muon spectra.

\subsection{Earth Matter Effects in Neutrino Events}
\label{sec:evnu}

\begin{figure}[t]
\centering
\includegraphics[width=0.495\textwidth]{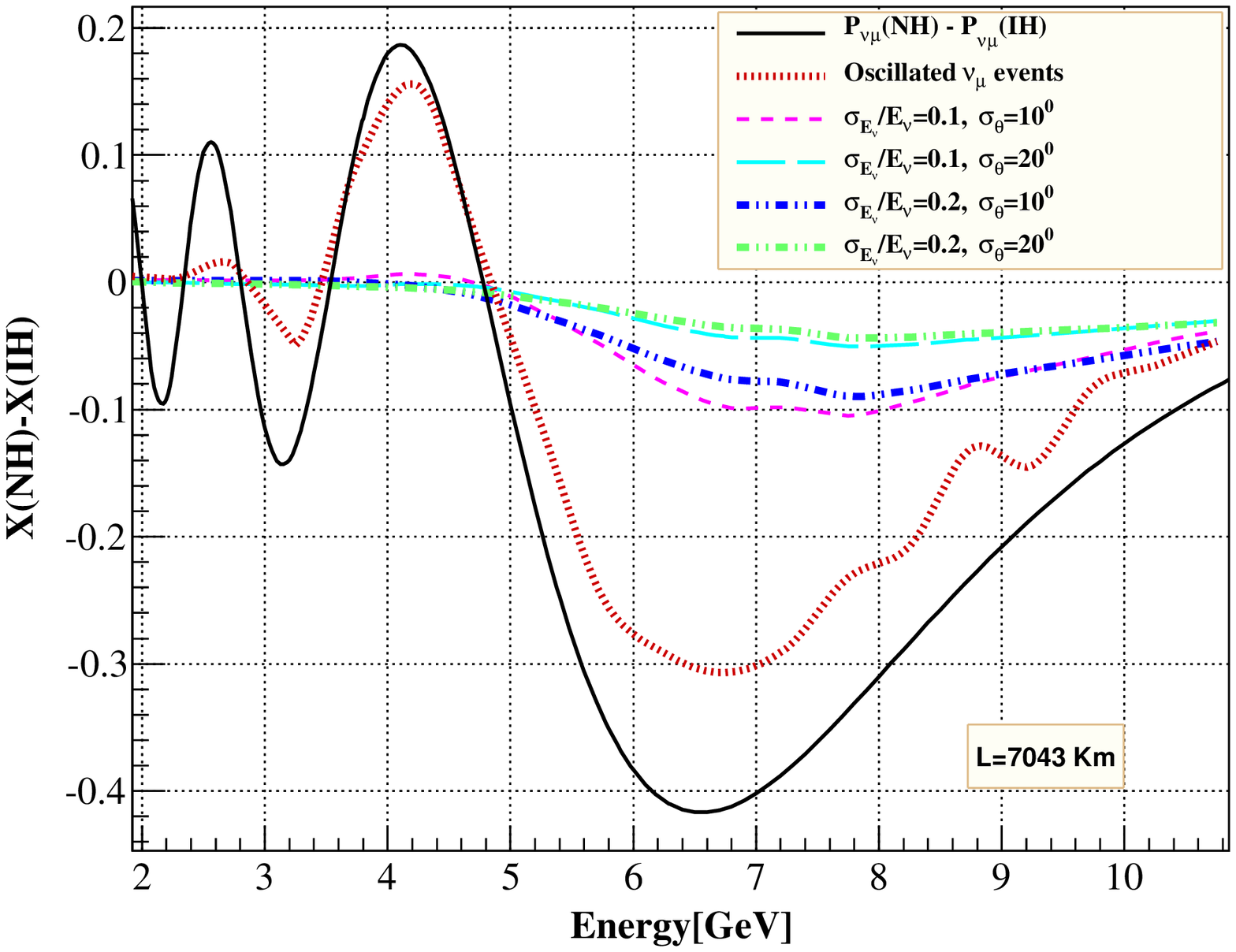}
\includegraphics[width=0.495\textwidth]{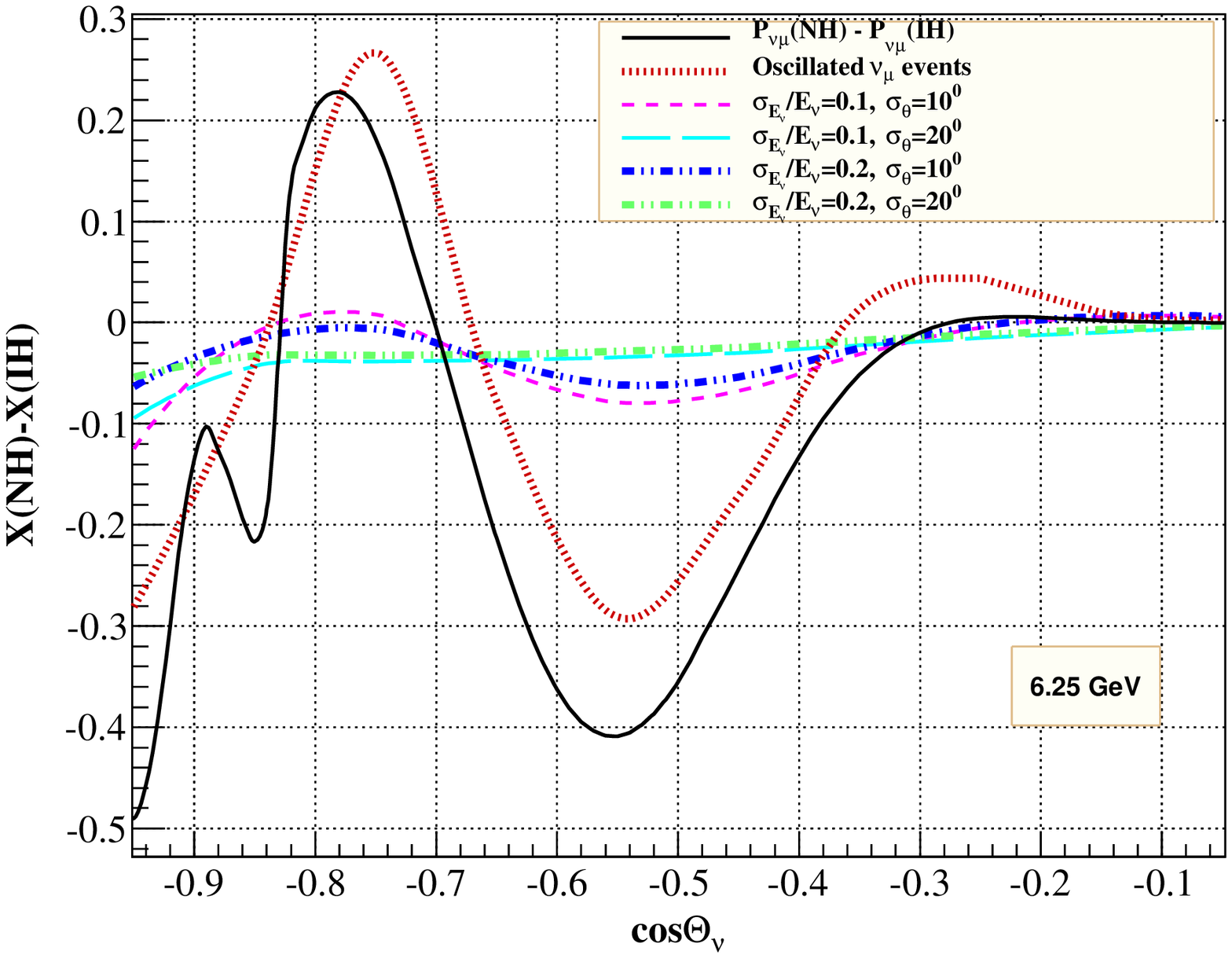}
\caption{The difference in the predicted hadron event rates for normal and 
inverted hierarchy, normalized to the no-oscillation hadron event rate, shown as 
a function of energy. The left panels shows the 
energy dependence of the hierarchy sensitivity, where 
the events are generated in the zenith angle 
bin marked by the path-length traversed at the mid-point of the bin. 
The right panel shows the corresponding plot as a function of the neutrino 
zenith angle in the energy bin corresponding to neutrino energy of 6.25 GeV.
The solid black lines 
show the $\Delta P_{\mu\mu}$ as in Fig. \ref{fig:evmuon}. 
The red dotted lines show the difference in the neutrino energy spectrum 
when no detector resolutions are added. The pink dashed, cyan long-dashed, 
blue dot-dashed and green dot-dashed lines show how the hierarchy sensitivity 
in the the neutrino spectrum changes as we include the detector resolutions in  
neutrino energy and neutrino zenith angle.
}
\label{fig:evnu}
\end{figure}

If the data is classified in terms of the neutrino energy and 
zenith angle, then the event spectrum is given as
\be
N_i^{\prime th} (\nu) =  {\cal N}
\sum_i
\int_{E_\nu^{min}}^{E_\nu^{max}}  \!\!\!dE_\nu^\prime
\int_{\cos\Theta_\nu^{min}}^{\cos\Theta_\nu^{max}}  \!\!\!d\cos\Theta_\nu^\prime
\,\, \,
R_{E_\nu}\,
R_{\Theta_\nu}\,
\bigg({\cal E}_i^\nu{\cal C}_i\,n_i(\nu) + {\overline {\cal E}_i^\nu}(1-\overline{{\cal C}_i})\,n_i(\nu)\bigg)\,
\,,
\label{eq:eventsthnu}
\ee
where all quantities are similar to those defined for the $\mu^-$ events in Eq. (\ref{eq:eventsthnu}) 
with the only difference that they now correspond to the neutrinos rather than to the muons. The 
charge identification efficiency is the only one which will be identical in the two cases. The resolution 
functions $R_{E_\nu}$ and $R_{\Theta_\nu}$
are also taken as Gaussian and their expression same as given for the muons but with the 
energy and angle resolution widths replaced by the corresponding ones for the neutrinos. 
A similar expression for the antineutrino events can also be written. 
\\

The 
difference in the neutrino event 
spectrum between the normal and inverted hierarchy normalized to the no-oscillation 
event rate 
is presented in Fig. \ref{fig:evnu}. The left panel of the figure shows the difference 
$\Delta X = X(NH)-X(IH)$ between the neutrino events for normal and inverted hierarchies 
as a function of the neutrino energy in the zenith angle bin corresponding to $\cos\Theta_\nu = -0.55$.
The right panel shows the $\Delta X$ as a function of the neutrino zenith angle for the energy 
bin corresponding to 6.25 GeV. The black lines in the figure show the $\Delta P_{\mu\mu}$, as before. 
The red dotted line is the observed 
neutrino event spectrum obtained when we take both the disappearance 
($P_{\mu\mu}$) as well as appearance ($P_{e\mu}$) channels into account. 
Note that as in the case of the muons, the inclusion of the appearance channel 
spoils the extent of hierarchy sensitivity in the neutrino spectrum. 
Yet, it is evident from a comparison of the Figs. \ref{fig:evmuon} and \ref{fig:evnu} that 
the net hierarchy sensitivity at this stage is greater in the neutrino case. The reason is because 
in the case of the muons there was a smearing of the spectral shape due to the 
kinematic averaging of the oscillation effects due to the interaction cross-section. 
However, the muon events have an advantage that the zenith angle of the muons 
can be very accurately reconstructed. Hence any further deterioration in the sensitivity comes  
mainly from the energy resolution of the muons. In the case of the neutrinos, both the energy as well as 
the zenith angle dependence of the signal gets smeared due to the finite detector resolutions. 
We show the impact of the resolution functions for four benchmark cases, 
$\sigma_{E_\nu}/E_\nu = 10\%,~\sigma_{\Theta_\nu}=10^\circ$ (pink dashed lines), 
$\sigma_{E_\nu}/E_\nu = 10\%,~\sigma_{\Theta_\nu}=20^\circ$ (cyan long-dashed lines), 
$\sigma_{E_\nu}/E_\nu = 20\%,~\sigma_{\Theta_\nu}=10^\circ$ (blue dot-dashed lines), and 
$\sigma_{E_\nu}/E_\nu = 20\%,~\sigma_{\Theta_\nu}=20^\circ$ (green dot-dashed lines). 
The figure shows that the impact of the resolution functions on the net hierarchy dependence is 
sharp. While both the energy as well as the angular resolutions result in reducing the hierarchy effect, 
the role of the angle resolution can be seen to be greater. It should be mentioned here that the 
benchmark resolution functions used here for illustration, range from being realistic to rather 
optimistic. Once we include these resolution functions, the $\Delta X \simeq -0.05$ to -0.1, depending 
on the values of the widths of the resolution functions. Comparison with the Fig. \ref{fig:evmuon} 
reveals that the $\Delta X$ which we are getting with the neutrino spectrum are in general 
of the same order or even lower than what we had obtained using the events binned in muon energy and 
zenith angle. Therefore, as we will see also from the $\chi^2$ analysis later, that the analysis of the 
neutrino data fails to give a $\Delta \chi^2$ for the wrong hierarchy better than that obtained from the 
muon analysis, unless one assumes extremely optimistic resolution functions for the neutrino. 
\\

\section{Mass Hierarchy Sensitivity with only Muon Events}
\label{sec:muonchi}

We begin with the analysis of only the muon event sample of the experiment. 
The mass hierarchy sensitivity of ICAL@INO using muon events was studied in detail by the 
INO collaboration and reported in \cite{mh}. The detector specification used in that 
work was the ones obtained from detailed simulations of the ICAL detector using Geant-based 
detector simulation codes \cite{inomuon}. These are however the first set of simulation results 
coming from dedicated studies of the ICAL detector performance. 
The study of the ICAL detector response to muons is still on-going. Therefore, 
it is expected that these results 
will be further refined and improved upon. Hence it is pertinent at this point to study 
how much the mass hierarchy sensitivity of ICAL@INO or any other similar 
detector could be improve with the improvement in the detector response to muons. 
\\

The muon reconstruction efficiency, 
charge identification efficiency, energy and angle resolutions were seen to be 
a function of both the muon energy as well as zenith angle \cite{inomuon}. 
A snap-shot of the detector response to muons was shown Fig. 1 of \cite{mh} and 
Fig. 1 of \cite{precision}. One can see from these plots that the reconstruction 
efficiency of the muons is roughly in the ballpark of 80-90\% over most of the 
energy and zenith angle range of the muons, while the charge identification range 
is mostly around 99\%. The zenith angle resolution of the muons is seen to be 
around 0.01 in $\cos\Theta_\mu$ from these plots, while the  the energy resolution is seen to be 
around 10\%-15\%. While the zenith angle resolution of the detector 
found from the first set of simulations is seen to be extremely good, the 
energy resolution could be improved upon. 
\\

Since it appears that the energy resolution of the muons might have room for improvement 
in the future, in what follows we will study the improvement expected in the mass hierarchy 
sensitivity of magnetized detector such as ICAL@INO, 
as we improve the muon energy resolution. To clearly show the 
impact of better muon energy resolution on the mass hierarchy, we assume all other 
detector performance parameters fixed at a plausible value. In particular, we take 
muon charge identification 
efficiency ${\cal C} = 99$\% and muon zenith angle resolution in $\cos\Theta_\mu$ of 
$\sigma_{\Theta_\mu}=0.01$. 
These values more-or-less agree with the values obtained by the INO collaboration 
from detailed simulations, however we take them to be flat over all muon energy and zenith 
angle bins. The muon reconstruction efficiency could change depending on the cuts 
imposed on the data to improve the muon energy resolution. Hence, we also work with different 
muon reconstruction efficiency. 
The muon reconstruction efficiency and the muon energy resolution 
is also taken to be flat over all bins. We also take all detector specifications to be 
the same for $\mu^-$ and $\mu^+$ events.
\\

\begin{figure}[t]
\centering
\includegraphics[width=0.48\textwidth]{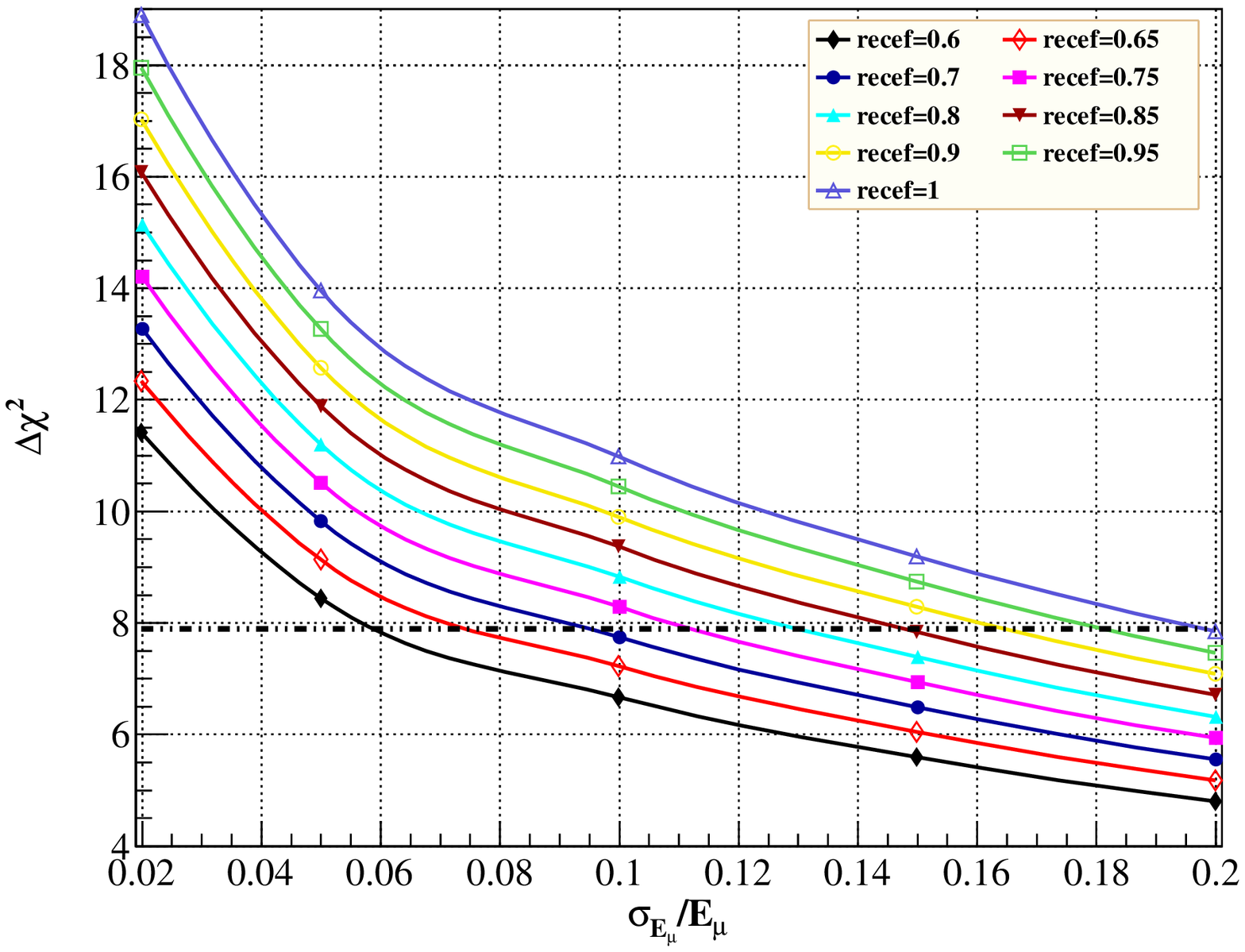}
\includegraphics[width=0.48\textwidth]{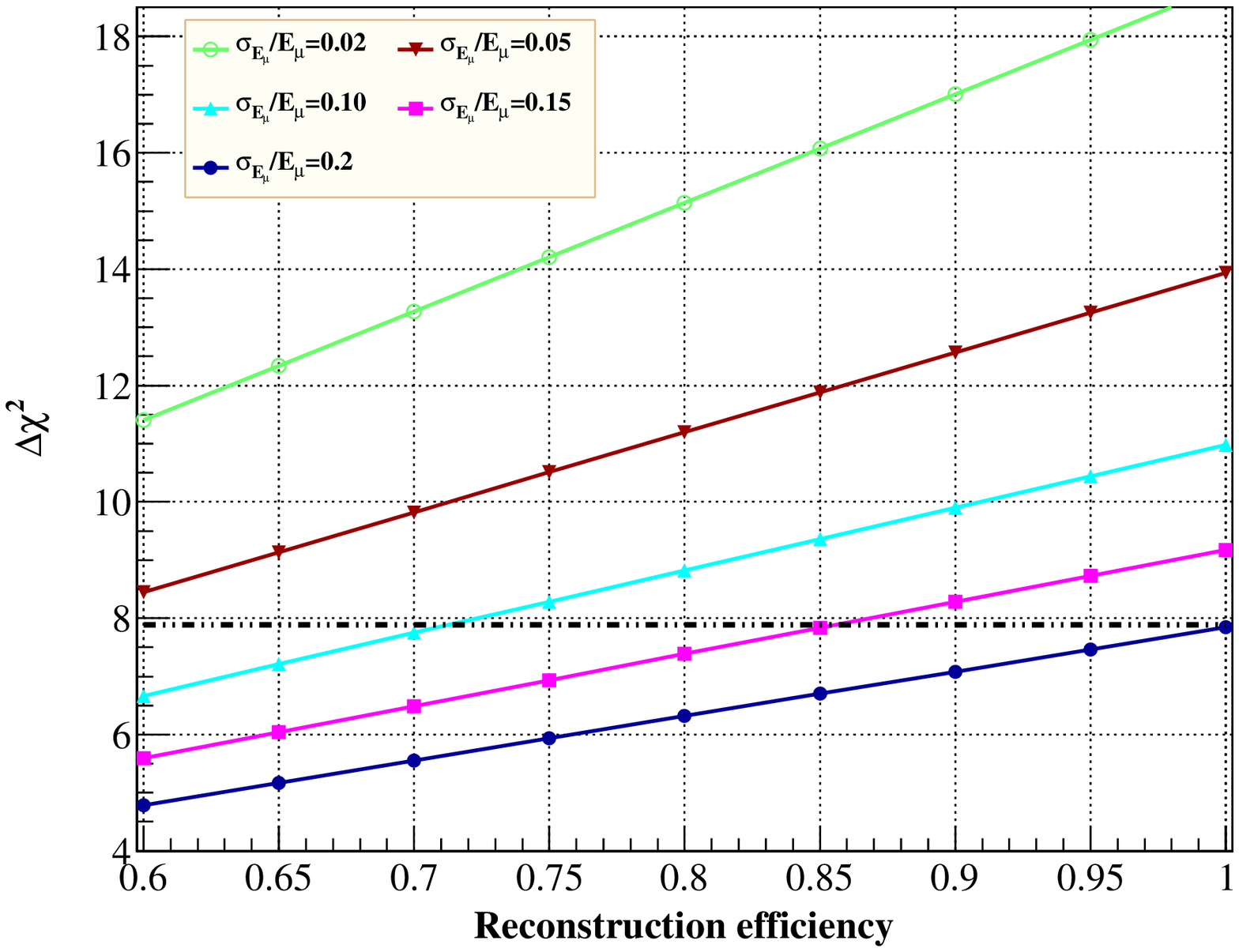}
\caption{The $\Delta \chi^2$ corresponding to the mass hierarchy sensitivity as a function of the
width of the muon resolution function (left panel) and muon reconstruction efficiency (right panel). 
We show this dependence for different fixed values of the reconstruction efficiency in the left panel and 
different fixed values of the width of the muon resolution function (right panel). The black dot-dashed 
lines in the figure show the $\Delta \chi^2$ obtained from the muon analysis performed in 
\cite{mh}, using the detector response to muons obtained by the INO collaboration.
}
\label{fig:muresol}
\end{figure}

To see how well the mass hierarchy can be determined at this experiment we simulate the 
data for normal mass hierarchy at the values of the oscillation parameters given in Table \ref{tab:param}. 
This data is then fitted with the wrong inverted hierarchy using a binned $\chi^2$ analysis. The 
$\chi^2$ function is defined for the $\mu^-$ sample as 
\be
\chi^2(\mu^-)={\min_{\{\xi_j\}}} \displaystyle\sum\limits_{i=1}^{N_{b}^\mu} \left
[2\left(N_i^{th}(\mu^-)-N_i^{ex}(\mu^-)\right) 
+2N_i^{ex}(\mu^-)ln\left(\frac{N_i^{ex}(\mu^-)}{N_i^{th}(\mu^-)}\right)\right] + \displaystyle\sum\limits_{j=1}^k \xi_j^{2}
\,,
\label{eq:chisqino}
\ee 
\be
N_i^{th}(\mu^-)=N_i^{\prime{th}}(\mu^-)\left(1+\displaystyle\sum\limits_{j=1}^k \pi_i^{j}{\xi_j}\right) + 
{\cal O}(\xi_k^2)
\,
\label{eq:evth}
\ee
where $N_i^{ex}(\mu^-)$ 
is the observed number of $\mu^-$ events in the $i^{th}$ bin 
and 
$N_i^{\prime{th}}(\mu^-)$ is the 
corresponding theoretically predicted event 
spectrum given by Eq. (\ref{eq:eventsth}). 
The 
 $\pi^j_i$ in Eq. (\ref{eq:evth}) 
 is the $j^{th}$ systematic uncertainty in the $i^{th}$ bin and 
$\xi_j$ is the pull variable corresponding to the uncertainty $\pi^j$. 
We have included five systematic uncertainties in our analysis. They are, an overall 
flux normalization error of 20\%,  a cross-section uncertainty of 10\%, a 5\% uncertainty on 
the zenith angle dependence of the fluxes,  
an energy dependent ``tilt factor", and a 5\% additional overall uncertainty.
A similar expression for the $\mu^+$ events can be written and the total 
$\chi^2$ is the sum of the two samples
\be
\chi^2(\mu) = \chi^2(\mu^-) + \chi^2(\mu^+)
\,.
\label{eq:chiino}
\ee
This $\chi^2(\mu)$ is minimized over the full set of pull variables $\{\xi_j\}$. 
Further details on the method of the $\chi^2$ analysis can be found in \cite{mh,precision}. 
All results presented in this paper are for $50\times 10$ kton-yrs of exposure of the experiment. 
The difference $\Delta \chi^2 = \chi^2(IH) - \chi^2(NH)$ is used as a measure of the 
mass hierarchy sensitivity of the experiment. 
\\

A discussion on the number of muon bins $N_b^\mu$ is in order. In \cite{mh} it was shown that 
for the excellent zenith angle resolution of the ICAL@INO detector, it makes sense to use 
80 $\cos\Theta_\mu^\prime$ bins of width 0.025, which is somewhat congruent with the 
width of the resolution function. In this work, since we have used a zenith 
angle resolution for the muons of 0.01 which is in the same ballpark as that obtained in the 
INO simulations, we continue to bin the muon zenith angle in 80 bins. Likewise, the 
energy bins should also correspond to the width of the energy resolution function. 
For the energy resolution obtained by the INO collaboration, 20 energy bins of width 0.5 GeV
was seen to be optimal in \cite{mh}. However, in this work we work with varying 
muon energy resolution. 
In particular, we vary $\sigma_{E_\mu}/E_\mu$ between 2\% and 20\%. 
In order to see the full impact of the improved energy resolution of 2\%, we 
bin the energy spectrum into 80 bins as well. We realize that binning the atmospheric 
neutrino events with $50\times 10$ kton-year exposure in $80\times 80$ bins will 
reduce the number of events drastically in each bin. However, first, we use the 
Poissonian definition for the $\chi^2$ function which can consistently calculate the 
$\chi^2$ even for very small event rates. Second, we take a purely  
phenomenological stance in this paper 
that if needed one could suitably increase the exposure of the experiment to 
compensate for the increased number of bins. And last but not the least, the 
experimental collaboration would perform an unbinned likelihood analysis of the data 
and in that case their statistical significance would roughly match with ours. 
Therefore, in this section we keep the number of energy bins fixed at 80 between 
1 GeV and 11 GeV. However, in the next section when we combine the $\chi^2$ 
corresponding to the hadron 
data with that of the muon data, then we present results for both 80 bins as well as 
20 bins between 1 GeV and 11 GeV.
\\

The dependence of the mass hierarchy sensitivity to the muon energy resolution and 
muon reconstruction efficiency are shown in Figs. \ref{fig:muresol}. 
In the left panel on 
Fig. \ref{fig:muresol} we show the $\Delta \chi^2$ as a function of the muon energy 
resolution $\sigma_{E_\mu}/E_\mu$, for different values of the muon reconstruction 
efficiency $\epsilon$. We vary $\sigma_{E_\mu}/E_\mu$ continuously between 20\% and 2\% and 
repeat this for $\epsilon = 70\%$ to $100\%$. 
For a given $\epsilon$, we see that the $\Delta \chi^2$ falls sharply as the 
$\sigma_{E_\mu}/E_\mu$ is increased from 2\% to about 6\%. Thereafter, the rate of fall of 
$\Delta \chi^2$ reduces, and it falls steadily as $\sigma_{E_\mu}/E_\mu$ worsens. 
As expected, $\Delta \chi^2$ is seen to increase with the reconstruction efficiency 
$\epsilon$. For comparison, we show by the black dot-dashed lines in the figure, the $\Delta \chi^2$ 
obtained in \cite{mh} using the full detector simulation results for the muon analysis. 
\\

In the right panel of Fig. \ref{fig:muresol} we show the $\Delta \chi^2$ 
as a function of the reconstruction 
efficiency $\epsilon$, for different values of the muon energy resolution $\sigma_{E_\mu}/E_\mu$. 
The $\Delta \chi^2$ is seen to increase linearly with $\epsilon$. This is not surprising as 
in the way we have included it in our analysis, 
$\epsilon$ linearly increases the statistics of the experiment. Since the mass hierarchy 
signal in an ICAL@INO like experiment for a
$50\times 10$ kton-yrs exposure is still in the statistics 
dominated regime, the $\Delta \chi^2$ grows linear with more data. 
\\

\begin{figure}[t]
\centering
\includegraphics[width=0.495\textwidth]{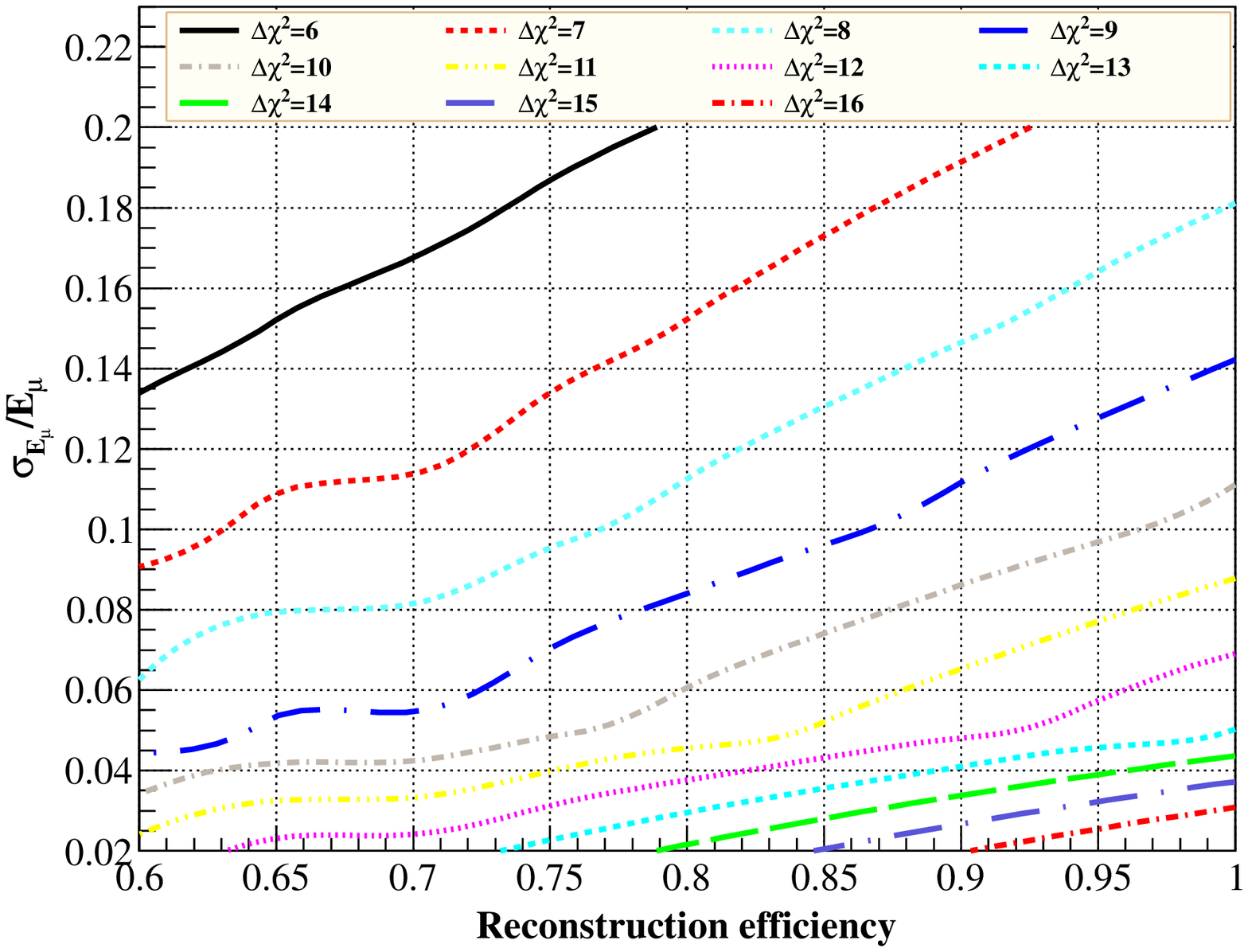}
\includegraphics[width=0.495\textwidth]{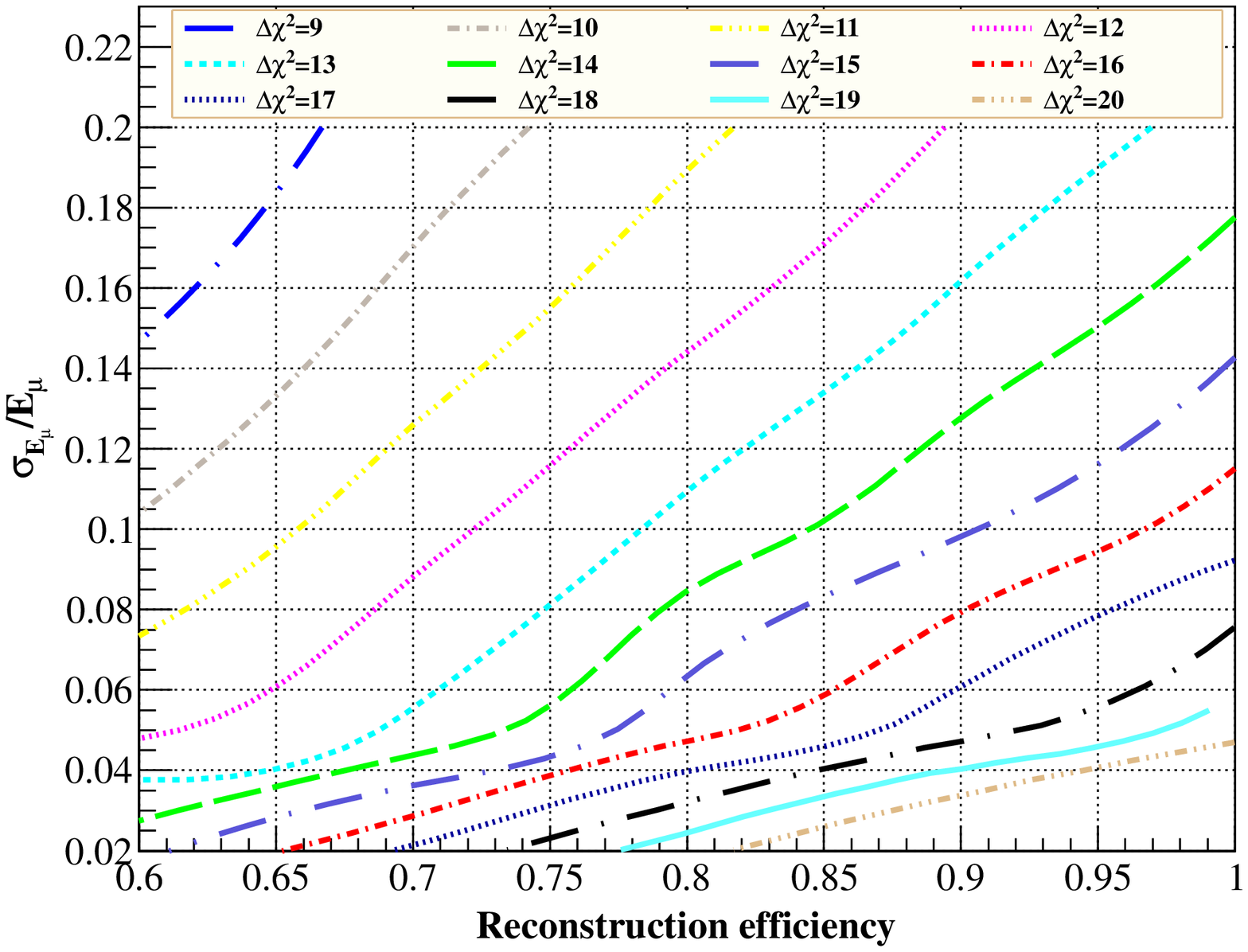}
\caption{Constant $\Delta \chi^2$ contours in the reconstruction efficiency and 
energy resolution plane. The left panels shows the contours from the analysis 
which uses only the muon data from the experiments. The right panels shows the 
contours for the analysis in which both the muon and hadron data were included in the 
combined statistical analysis. 
}
\label{fig:cont}
\end{figure}

Finally, in the left panel of Fig. \ref{fig:cont}, 
we show contours of constant $\Delta \chi^2$ obtained in the reconstruction efficiency and 
energy resolution plane, from the analysis of the 
muon events in the detector. We show the contours for 
$\Delta \chi^2 = 6$ to 16. The figure shows that for energy resolution in the range of 2\% to 3\% 
and reconstruction efficiency above 90\%, one would get a $4\sigma$ signal for the 
neutrino mass hierarchy from the analysis of the muon data alone. We reiterate the 
the reconstruction efficiency can be compensated by adjusting the exposure of the 
experiment. On the other hand, a $3\sigma$ measurement of the mass hierarchy 
from the muon analysis alone seems to be extremely plausible for reasonable 
range of values for the muon energy resolution and reconstruction efficiency. 
\\

\section{Mass Hierarchy Sensitivity with Muon Plus Hadron Events}

We had seen in Fig. \ref{fig:evhadron} that the hadron events also 
exhibit a difference between the event rates expected for the normal and 
inverted hierarchy and hence carry 
mass hierarchy sensitivity to some extent. In this section, we quantify the 
improvement in the sensitivity expected in the mass hierarchy sensitivity  
of magnetized iron detectors if the hadron events were also added in the 
statistical analysis of the data. The total $\chi^2$ function associated with the 
$\mu^-$ is then given by 
\be
\chi^2(\mu^-+ hadron^-)&=& {\min_{\{\xi_j\}}}\Bigg [ ~\displaystyle\sum\limits_{i=1}^{N_{b}^\mu} \left
\{2\left(N_i^{th}(\mu^-)-N_i^{ex}(\mu^-)\right) 
+2N_i^{ex}(\mu^-)\ln\left(\frac{N_i^{ex}(\mu^-)}{N_i^{th}(\mu^-)}\right)\right \}  \\ \nonumber
&+& 
\displaystyle\sum\limits_{i=1}^{N_{b}^{H}} \left
\{2\left(H_i^{th}(\mu^-)-H_i^{ex}(\mu^-)\right) 
+2H_i^{ex}(\mu^-)\ln\left(\frac{H_i^{ex}(\mu^-)}{H_i^{th}(\mu^-)}\right)\right\} ~\Bigg ] + 
\displaystyle\sum\limits_{j=1}^k \xi_j^{2}
\,,
\label{eq:chitot}
\ee
where $N_b^\mu$ and $N_b^H$ are the total number of $\mu^-$ and hadron bins, respectively,
$N_i^{ex}(\mu^-)$ and $H_i^{ex}(\mu^-)$ are the data in the $\mu^-$ and the associated 
hadrons, respectively, 
while the theoretical predictions are given as
\be
H_i^{th}(\mu^-)=H_i^{\prime{th}}(\mu^-)\left(1+\displaystyle\sum\limits_{j=1}^k \eta_i^{j}{\xi_j}\right) + 
{\cal O}(\xi_k^2)
\,,
\label{eq:evthmu}
\ee
where $H_i^{\prime{th}}(\mu^-)$ is given by Eq. (\ref{eq:evhadron}) and 
all other quantities are as defined in the previous section. 
Since the hadrons are tagged with their corresponding muon 
from the respective charged current interaction, we can distinguish between the hadrons 
associated with the $\mu^-$ from the ones associated with the $\mu^+$. 
The $-$ sign as 
superscript in $hadron^-$  in Eq. (\ref{eq:chitot}) signifies that the hadrons implied there 
are ones associated with the $\mu^-$ events.  
Therefore, the 
combined $\chi^2$ using all data is given by
\be
\Delta \chi^2(\mu + hadron) = \Delta \chi^2(\mu^- + hadron^- ) + \Delta \chi^2(\mu^++hadron^+) 
\,.
\ee
The hadron data is binned in the way described earlier in connection with the Fig. \ref{fig:evhadron}. 
We first present the results of the statistical analysis where the hadrons are binned in 
energy only (cf. Eq. (\ref{eq:evhadron}). We then introduce the zenith angle resolution 
function of the hadron as well and further bins the data into hadron angles.

\subsection{Hadron Data Binned in Energy Only}

\begin{table}[h]
\begin{center}
\begin{tabular}{|c|c|c|c|c|c|c|}
\hline
$N_{bin}^{E_\mu}$&$\sigma_{E_\mu}/E_\mu$& Muon  & $\Delta \chi^2(\mu)$ & \multicolumn{3}{|c|}  {$\Delta \chi^2(\mu+hadron)$} 
\\ \cline{5-7}
&&Rec Eff &  &$\frac{\sigma_H}{E} = \frac{0.8}{\sqrt{E}} + 0.2$ 
& $\frac{\sigma_H}{E} = \frac{0.68}{\sqrt{E}} + 0.02$ & $\frac{\sigma_H}{E} = \frac{0.56}{\sqrt{E}} + 0.02$
 \\ \hline \hline
20 & 0.12&70\% & 7.0 & 9.0 & 9.5 & 9.7 \\ \hline
20 & 0.12&80\% & 8.0 & 10.3 & 10.8 & 11.0 \\ \hline
20 & 0.12&90\% & 9.0 & 11.6 & 12.1 & 12.4 \\ \hline \hline
20 & 0.02 & 70\% & 10.2& 12.2 & 12.6 & 12.8 \\ \hline
20 & 0.02 & 80\% & 11.6&  13.9 & 14.4 & 14.6 \\ \hline
20 & 0.02 & 90\% & 13.0& 15.6 & 16.2 & 16.4 \\\hline \hline
80 & 0.02 & 70\% & 13.3& 15.3 & 15.7 & 15.9 \\ \hline
80 & 0.02 & 80\% & 15.1&  17.5 & 18.0 & 18.2 \\ \hline
80 & 0.02 & 90\% & 17.0& 19.6 & 20.2 & 20.4 \\\hline
\end{tabular}
\end{center}
\caption{\label{tab:chihad1}
The $\Delta \chi^2$ obtained for ruling out the wrong mass hierarchy obtained by 
combining the hadron data with the muon data, where we bin the hadron data in 
10 equal hadron energy bins between $E_H = 0.5$ GeV and 10.5 GeV (1 GeV bins). 
The muons are binned in 80 zenith angle bins of 
width $\Delta \cos\Theta_\mu = 0.025$, while the energy bins for the muons are varied, 
and shown in the first column. The second column shows the muon 
energy resolution, third column gives the 
reconstruction 
efficiency, the fourth column the $\Delta \chi^2$ obtained using only the muon data, 
while the fifth, sixth and seventh columns give the $\Delta \chi^2$ from the 
combined analysis of the muon and the hadron data. 
}
\end{table}

We present in Table \ref{tab:chihad1} the improvement in $\Delta \chi^2$ we obtain by 
including the hadron energy binned data into our $\chi^2$ analysis, where we bin the hadron data in 
10 equal hadron energy bins between $E_H = 0.5$ GeV and 10.5 GeV (1 GeV bins). 
The combined $\Delta \chi^2(\mu + hadron)$ is shown 
in Table \ref{tab:chihad1} for different cases of 
muon energy resolutions and reconstruction efficiencies, as well for different hadron energy resolutions. 
We show the sensitivity for three benchmark values of the muon reconstruction efficiency of 
70\%, 80\% and 90\%. The current ICAL@INO reconstruction efficiency corresponds to roughly 80-90\%. 
The muon energy resolution is varied between 12\%, the value obtained in the current simulations 
being performed by the INO collaboration \cite{inomuon}, and 2\%, which is the projected ambitious goal 
for this detector parameter. For 12\% energy resolution we work with 20 muon energy bins between 
muon energy of 1 GeV and 11 GeV, which corresponds well with this value of the energy resolution. 
However, for the 2\% energy resolution case we show the results for two choices of 
energy bins -- 20 muon energy bins and 80 muon energy bins 
between muon energies of 1 GeV and 11 GeV. The choice of $N_{bin}^{E_\mu}=80$ is 
commensurate with the very good energy resolution of 2\% and should be the logical one 
to be used. However, we 
also show the results for the more conservative choice of $N_{bin}^{E_\mu}=20$. 
For the hadron energy resolution we repeat our analysis for three benchmark choices :-
\begin{enumerate}

\item $\frac{\sigma_{E_H}}{E_H} = \frac{0.8}{\sqrt{E_H(GeV)}} + 0.2$. This is 
roughly the kind of hadron energy resolution that has been obtained by the 
INO collaboration through the Geant-based simulation of their detector response to 
hadrons \cite{inohadron}.

\item $\frac{\sigma_{E_H}}{E_H} = \frac{0.68}{\sqrt{E_H(GeV)}} + 0.02$. This is the 
the hadron energy resolution obtained by testing an 8 ton prototype magnetized 
iron calorimetric detector equipped with 23 m$^2$ of glass Resistive Plate Chambers (RPC) 
with the T7-PS beam at CERN by the 
the MONOLITH collaboration \cite{monolithenergy}. 

\item $\frac{\sigma_{E_H}}{E_H} = \frac{0.56}{\sqrt{E_H(GeV)}} + 0.02$. This is the 
hadron energy resolution reported by the MINOS collaboration.

\end{enumerate}
The results for these three different $\sigma_{E_H}/E_H$ are presented in the last three columns of 
Table \ref{tab:chihad1}.  A comparison of the $\Delta \chi^2(\mu)$ obtained from 
using only the muon data in the analysis (given in the fourth 
column of Table \ref{tab:chihad1}) with the $\Delta \chi^2(\mu+hadron)$ obtained by 
adding the corresponding hadron data to the muon data 
(given in the last three columns of Table \ref{tab:chihad1}) 
shows the increase in the mass hierarchy sensitivity of the experiment 
due to the contribution of the hadron data. For instance, we can see that with 
80\% muon reconstruction efficiency and 12\% energy resolution, 
one gets $\Delta \chi^2(\mu)=8.0$. On adding the hadron data this 
increases to $\Delta \chi^2(\mu+hadron)=10.3-11.0$, depending on the 
hadron energy resolution of the detector. Therefore, we get a contribution of about 
2-3 to the $\Delta \chi^2$ from just the hadron events in the detector. Since the hadron 
events are tagged with the muon events, an increase in the muon reconstruction efficiency 
brings about a corresponding increase in the statistics of the hadron events and hence 
an increase in the contribution of the hadron events to the mass hierarchy sensitivity. 
For 90\% reconstruction efficiency, we see that hadron data gives a contribution 
of about 2.5-3.5 to the $\Delta \chi^2$. 
For 80\% muon reconstruction efficiency, if we improve the 
muon energy resolution $\sigma_{E_\mu}/E_\mu$ to 2\%, the corresponding $\Delta \chi^2(\mu)$ 
increases to 11.6 for $N_{b}^{E_\mu}=20$ bins and further to 15.1 for $N_{b}^{E_\mu}=80$.
On addition of the hadron events to the muon sample, the corresponding sensitivity reach 
numbers stand as 13.9-14.6 and 17.5-18.2, respectively. An increase of muon reconstruction 
efficiency to 90\% will give $\Delta \chi^2(\mu)=17$ and 
$\Delta \chi^2(\mu + hadron)=19.6-20.4$ in the most 
optimistic cases of the muon resolutions of 2\% with 80 muon energy bins. 
On the other hand, if we compare the increase in the $\Delta \chi^2$ as we 
change the hadron energy resolution $\sigma_{E_H}/E_H$, we find that the 
change is marginal. Indeed the Table \ref{tab:chihad1} shows that in going from 
$\sigma_{E_H}/E_H = 0.8/sqrt{E_H} + 0.2$ to $\sigma_{E_H}/E_H = 0.56/sqrt{E_H} + 0.02$
the $\Delta \chi^2(\mu + hadron)$ changes by less than $\sim 1$. 
\\

\subsection{Hadron Data Binned in Energy and Zenith Angle}
\label{sec:chihad2}

\begin{table}[h]
\begin{center}
\begin{tabular}{|c|c|c|c|c|c|c|}
\hline
$N_{bin}^{E_\mu}$&$\sigma_{E_\mu}/E_\mu$& $\sigma_{\Theta_H} $ & $\Delta \chi^2(\mu)$ & \multicolumn{3}{|c|} {$\Delta \chi^2(\mu+hadron)$} 
\\ \cline{5-7}
&&&  &$\frac{\sigma_{E_H}}{E} = \frac{0.8}{\sqrt{E}} + 0.2$ 
& $\frac{\sigma_{E_H}}{E} = \frac{0.68}{\sqrt{E}} + 0.02$ & $\frac{\sigma_{E_H}}{E} = \frac{0.56}{\sqrt{E}} + 0.02$
 \\ \hline \hline
20 & 0.12& $ \frac{16.67}{\sqrt{E}} + \frac{12.12}{E} $& 8.0 & 11.3 & 12.1 & 12.5 \\ \hline
20 & 0.12& $ \frac{10.4}{\sqrt{E}} + \frac{10.1}{E} $& 8.0 & 11.6 & 12.5 & 12.9 \\ \hline \hline
20 & 0.02 &  $ \frac{16.67}{\sqrt{E}} + \frac{12.12}{E} $&11.6 & 14.9 & 15.7 & 15.4 \\ \hline
20 & 0.02 & $ \frac{10.4}{\sqrt{E}} + \frac{10.1}{E} $ & 11.6& 15.2 & 16.1 & 16.6 \\\hline \hline
80 & 0.02 & $ \frac{16.67}{\sqrt{E}} + \frac{12.12}{E} $ &15.1 & 18.8 & 19.3 & 18.9 \\ \hline
80 & 0.02 & $ \frac{10.4}{\sqrt{E}} + \frac{10.1}{E} $ & 15.1& 18.5 & 19.7 & 20.12 \\\hline
\end{tabular}
\end{center}
\caption{\label{tab:chihad2}
The $\Delta \chi^2$ obtained for ruling out the wrong mass hierarchy obtained by 
combining the hadron data with the muon data, where we bin the hadron data in 
five equal hadron energy bins between $E_H = 0.5$ GeV and 10.5 GeV (2 GeV bins) 
and 5 equal hadron zenith angle bins between $\Theta_H = 0^\circ$ and 
$180^\circ$ ($36^\circ$ bins). The muons are binned in 80 zenith angle bins of 
width $\Delta \cos\Theta_\mu = 0.025$, while the energy bins for the muons are varied, 
and shown in the first column. We have take the reconstruction efficiency as 80\% for 
all cases in this Table. 
The second column shows the muon 
energy resolution, the third column the $\Delta \chi^2$ obtained using only the muon data, 
while the fourth, fifth and sixth columns give the $\Delta \chi^2$ from the 
combined analysis of the muon and the hadron data. 
}
\end{table}

The magnetized iron calorimeter is capable of measuring the angle of the hadron shower. 
In order to use the full energy and angle information on the hadrons in the analysis, we 
bin the data in both hadron energy as well as zenith angle. For every muon 
zenith angle bin, we bin the hadrons into 
5 energy bins between $E_H = 0.5$ and 10.5 GeV of bin width 2 GeV and 
5 zenith angle bins of bin width $36^\circ$. We smear 
the zenith angle binned data by the Gaussian smearing function $R_{\Theta_H}$ in addition 
to the smearing functions for the muon zenith angle resolution and hadron energy resolution 
shown in Eq. (\ref{eq:evhadron}).
The simulation results for the angular resolution of the hadron shower in ICAL are yet to be made 
public by the INO collaboration. Therefore, in what follows we will use the following two 
cases for the zenith angle resolution for the hadron shower:
\begin{enumerate}
\item $\sigma_{\Theta_H} = \frac{16.67}{\sqrt{E_H}} + \frac{12.12}{E_H}$, taken from the 
MINOS proposal.
\item $\sigma_{\Theta_H} = \frac{10.4}{\sqrt{E_H}} + \frac{10.1}{E_H}$, taken from the 
analysis of the MONOLITH prototype data in the T7-PS beam 
at CERN \cite{monolithangle}.
\end{enumerate}
The results obtained by including the full energy as well as angle binned hadron 
data along with the muon data in the statistical analysis, are shown in Table \ref{tab:chihad2}.
We find that the binning in hadron zenith angle brings in a further mild increase in the total 
$\Delta \chi^2$. The increase in the $\Delta \chi^2$ due to the introduction of the 
hadron events into the analysis is seen to be improving it by $\simeq 5$. 
With a 80\% reconstruction efficiency and muon energy resolution of 
2\% one could get a total $\Delta \chi^2\simeq 20$ (16) if one choose to work with 
80 (20) muon energy bins.
\\

The constant $\Delta \chi^2$ contours for the full data set in the 
reconstruction efficiency and 
muon energy resolution plane is shown in the right panel of Fig. \ref{fig:cont}. 
Here we have included the energy and zenith angle binned hadron data into the 
analysis with $\sigma_{E_H}/E_H = 0.68/\sqrt{E_H} + 0.02$ \cite{monolithenergy} and 
$\sigma_\Theta = 10.4/\sqrt{E_H} + 10.1/{E_H}$ \cite{monolithangle}. The contours are 
shown 
$\Delta \chi^2 = 9$ to 20. 
We can see that getting a $3\sigma$ signal for the mass hierarchy becomes 
very easy for most values of the reconstruction efficiency and muon energy resolution 
once the hadron data is added. For muon energy resolution in the range of 
2\% to 10\% and reconstruction efficiency greater than 65\%, we should get 
$4\sigma$ sensitivity to the mass hierarchy. 
If the muon energy resolution is in the range of 2\% to 4\% 
and reconstruction efficiency above 80\%, one would get a $4.5\sigma$ signal for the 
neutrino mass hierarchy from the analysis of the combined data. 
\\

\section{Mass Hierarchy Sensitivity with Neutrino Events}
\label{sec:nuchi}

\begin{figure}[t]
\centering
\includegraphics[width=0.58\textwidth]{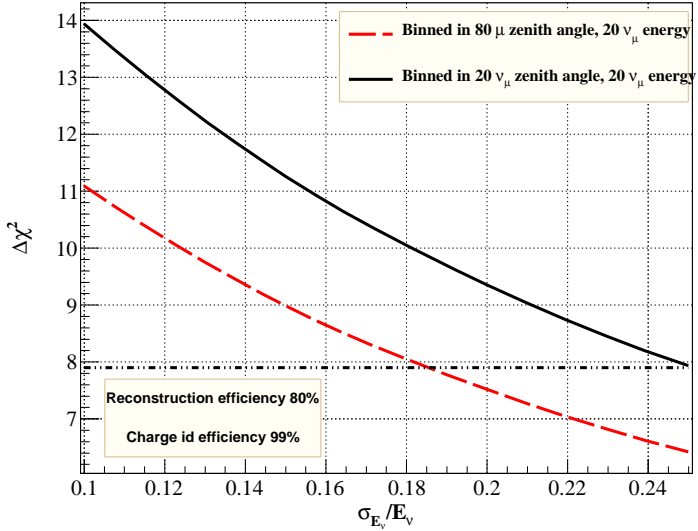}
\caption{The $\Delta \chi^2$ corresponding to the mass hierarchy sensitivity as a function of the
width of the neutrino resolution function. The black line shows the expected sensitivity 
when we use the binning in neutrino zenith angles with the neutrino zenith angle resolution 
function taken from the MONOLITH proposal \cite{monolithproposal}. The red dashed line 
shows the expected sensitivity 
when we use the binning in muon zenith angles with the muon zenith angle resolution 
width of 0.01 in $\cos\Theta_\mu$. 
The black dot-dashed 
lines in the figure show the $\Delta \chi^2$ obtained from the muon analysis performed in 
\cite{mh}, using the detector response to muons obtained by the INO collaboration.
}
\label{fig:nuchi}
\end{figure}

As discussed before, the measured energy and angle of the muon, as well as the 
energy and angle of the hadron, can be combined to reconstruct the 
energy and zenith angle of the neutrino. The width of the  corresponding energy and 
angle resolution was seen to be crucial when we presented the neutrino event 
spectrum in section \ref{sec:evnu}.  Here we show the statistical significance 
with which the wrong hierarchy can be ruled out by the neutrino analysis. The 
$\chi^2$ is defined similar to what we had for the muon analysis (cf. 
Eq. (\ref{eq:chisqino})), with just the 
muon bins replaced by the neutrino bins. We show the 
results of our analysis in Fig. \ref{fig:nuchi}. The black solid line shows the 
$\Delta \chi^2$ obtained for the wrong hierarchy when we bin the data in 
20 neutrino energy bins between 1 GeV and 11 GeV, and 20 zenith angle bins 
between $-1$ and $+1$. We have taken the 
neutrino reconstruction efficiency of 80\% and 
charge identification efficiency of 99\% in this figure. 
The neutrino energy resolution $\sigma_{E_\nu}/E_\nu$ is 
taken as flat over all bins and is varied in the x-axis. The neutrino zenith angle 
resolution adopted in this figure has been taken from Fig. 4.7 of the MONOLITH 
proposal \cite{monolithproposal}. While the average of neutrino zenith angle 
resolution is around $11^\circ$, in the energy range of interest, the zenith angle 
resolution is seen to be around $5-7^\circ$. 
The neutrino energy resolution in this proposal is 
quoted as being around 20\%. We can see from the figure that at 20\% energy 
resolution $\Delta \chi^2 \simeq 9.4$ from this analysis. \footnote{For a flat 10\% 
neutrino energy 
resolution and $10^\circ$ zenith angle resolution function, we obtain 
$\Delta \chi^2 = 10.4$ from the neutrino analysis.} 
This is comparable to the sensitivity obtained 
with just the muon data with 80\% efficiency and energy resolution of 12\%. 
If the hadron data is added to the muon data, then even with these modest 
detector response for the muons, we would get $\Delta  \chi^2 \simeq 12-13$ (cf. Table \ref{tab:chihad2}). 
If the muon energy resolution was improved to 2\%, then we would have a more than $4\sigma$ 
sensitivity to the mass hierarchy from the muon-plus-hadron analysis. Whereas, from the 
Fig. \ref{fig:evnu} we can see that the sensitivity from the neutrino analysis can 
never match these numbers, even for extremely optimistic energy resolution of 10\%. 
The reason can be traced to the event rates plots shown in Figs. \ref{fig:evmuon}, \ref{fig:evhadron} and 
\ref{fig:evnu}. The effect of the detector resolutions on the neutrino spectrum make it 
comparable, and sometimes even worse than the muon spectrum. If we add the hadron 
spectral data to the muon spectral analysis, we get an additional contribution to the 
hierarchy sensitivity which cannot be matched by the neutrino analysis. 
\\

For comparison, in Fig. \ref{fig:nuchi}
we also show the $\Delta \chi^2$ obtained
when we bin the data in neutrino energy and muon zenith angle bins. 
We keep the reconstruction efficiency at 80\% and charge identification efficiency at 99\%. 
The muon zenith angle resolution function is taken with width 0.01. We see that the 
$\Delta \chi^2$ obtained in this case is even worse than what we had obtained with the 
neutrino zenith angle analysis using the MONOLITH zenith angle resolution function. 
\\

\section{Conclusions}

Measurement of the neutrino mass hierarchy is the next important thing to do in neutrino physics. 
The atmospheric neutrino experiments could play a crucial role in this respect, by observing the 
hierarchy dependent earth matter effects. In this paper we consider the magnetized iron calorimeter 
as the atmospheric neutrino detector, such as the ICAL@INO. This class of detectors are sensitive 
to mainly muons, but have good energy and angle resolution and very good charge 
identification capabilities. In this paper we optimized the mass hierarchy sensitivity of the experiment 
with respect to the detector response functions and showed how using the hadron data as an 
independent input along with the muon data is most likely to give the best sensitivity to the 
neutrino mass hierarchy. This way of treating the data from atmospheric neutrinos, to 
the best of our knowledge, is being proposed for the first time in this paper. 
\\

We began with first discussing the earth matter effects 
in the muon neutrino survival probabilities which leads 
to the neutrino mass hierarchy sensitivity in this kind of 
experiment. We next showed how this sensitivity gets watered down 
when we consider the event rates in the detector  
(i) due to the inclusion of the appearance 
channel, (ii) due to the averaging effect of the charged current interaction which 
produces muons with any energy and angle allowed by the process, and 
(iii) with inclusion of detector resolutions. 
We showed how the mass hierarchy effect in the muon event sample, 
hadron event sample as well 
as the neutrino event sample decreases as we include these factors one by one. 
In the case of the muon and hadron spectrum, the major smearing of the 
earth matter effects come from the cross-section effect. The  
energy resolution brings about a further reduction in the signal,
however the effect is mild. 
We showed that despite this reduction in the earth matter effects, 
both the muons as well as the hadrons 
event spectra have mass hierarchy 
sensitivity, which survives even after including all the resolution functions. 
The neutrino analysis is affected only due to the appearance channel 
and the detector resolutions. However, the neutrino events turned out 
to be very sensitive to the detector 
energy and zenith angle resolution. It was shown that the hierarchy effects 
reduce sharply as we switch on the neutrino energy and angle resolutions, 
such that the net earth matter effect present in the neutrino spectrum 
becomes comparable to that in the muon spectrum. 
\\

We performed a $\chi^2$ analysis to quantify the reach of the experiment to measuring the neutrino mass 
hierarchy. We showed results for three different analysis.
\begin{itemize}

\item
We started with the analysis of only the muon events using a treatment similar to 
that in \cite{mh} but with flat efficiencies and resolution functions. 
Since the zenith angle resolution given by the INO collaboration is already 
very good, we fixed the zenith angle resolution for the muons at $\sigma_{\Theta_\mu} = 0.01$ in 
$\cos\Theta_\mu$ and showed how the mass hierarchy sensitivity could be improved by 
improving the muon energy resolution and muon reconstruction efficiency. If the muon 
energy resolution could be improved to 2\%, we could get a more than $4\sigma$ 
measurement of the mass hierarchy from the $50\times 10$ kton-year of muon data alone. 

\item 
We next included the hadron events as an additional input in the analysis along with the 
muon data. Since the detector is not expected to measure the hadron energy and angle as 
well as it can do for the muon and since we wanted to keep track of the particles coming from 
a given neutrino energy and angle, we tagged the hadron with their corresponding muon 
from the charged current interaction. Since the muon zenith angle is the best measured 
quantity we collect all hadrons in a given muon zenith angle bin. These hadrons are then 
binned in their energy and zenith angle. We defined a $\chi^2$ function for the combined 
analysis of the hadron and muon events with the so-binned muon and hadron data. 
The results showed that the hadron events bring in a noticeable improvement in the 
final sensitivity of the experiment to the neutrino mass hierarchy by increasing the 
$\Delta \chi^2$ by up to 5. The combined muon and hadron analysis is projected to 
give a $4.5\sigma$ sensitivity from a $50\times 10$ kton-year exposure, if one could 
achieve 2\% energy resolution and 80\% reconstruction efficiency 
in the muons. 

\item Finally we showed the mass hierarchy sensitivity expected from the  
analysis of the data in terms of the neutrino energy and angle. For the 
zenith angle resolution obtained by the MONOLITH collaboration, we showed the 
$\Delta \chi^2$ expected from the neutrino analysis as a function of the 
neutrino energy resolution. For the 20\% energy resolution quoted by the 
MONOLITH collaboration, one would get a little over $3\sigma$ signal 
for the neutrino mass hierarchy, which is lower than 
what we got from the combined muon and hadron analysis. 
We argued that this happens because the neutrino channel is 
very sensitive to the detector resolution functions. Once the detector 
resolution functions are imposed, the net earth matter effects in the neutrino 
channel becomes equal to, or sometimes even less than, the residual earth matter 
effects in the muon spectrum. When we add the hadron spectrum to the muon data, the 
total $\chi^2$ overshoots that expected from the neutrino analysis.

\end{itemize}

In conclusion, the neutrino mass hierarchy can be measured rather well from the 
observation of atmospheric neutrinos in magnetized iron calorimeters. 
The sensitivity can be significantly increased by improving the muon energy 
resolution of the detector. The addition of the hadron data into the analysis  
will improve the results even further, and return sensitivity reach 
which is better than what can be achieved from the neutrino analysis at these detectors.
For $\sin^22\theta_{13}=0.1$, $\sin^2\theta_{23}=0.5$,
a muon energy resolution of 2\%, reconstruction efficiency of 80\% and
exposure of $50\times 10$ kton-year,
we could get up to $4.5\sigma$ signal for the mass hierarchy from combining the muon
and hadron data. The signal will go up when the atmospheric data is combined with data
from other existing experiments, particularly NO$\nu$A.

\vskip 1cm
\noindent
{\Large \bf Acknowledgements}
\\

We thank Tarak Thakore and Moon Moon Devi for discussions and the INO 
collaboration for their continued support. 
S.C. acknowledges partial support from the European Union FP7 ITN INVISIBLES
(Marie Curie Actions, PITN-GA-2011-289442).


\end{document}